\documentclass[preprint,11pt]{elsarticle}
\usepackage{amsfonts,amscd,amsmath, amssymb,graphicx,color}

\usepackage{amssymb}
\newcommand{\normord}[1]{:\mathrel{#1}:}

\journal{Nuclear Physics B}

\begin{document}

\begin{frontmatter}



\title{\boldmath A Quantum Bound--State Description of Black Holes}


\author[a]{Stefan Hofmann}
\author[a,b]{and Tehseen Rug}

\address[a]{Arnold Sommerfeld Center for Theoretical Physics, 
	LMU-M\"unchen, Theresienstrasse 37, 80333 M\"unchen, Germany}
\address[b]{Max-Planck-Institut f\"ur Physik,
	F\"ohringer Ring 6, 80805 M\"unchen, Germany}

\begin{abstract}
A relativistic framework for the description of bound
states consisting of a large number of quantum constituents is presented,
and applied to black-hole interiors. At the parton level, the constituent distribution, 
number and energy density inside black holes are calculated, and gauge corrections 
are discussed. A simple scaling relation between the black-hole mass and constituent 
number is established.

\end{abstract}


\end{frontmatter}
\section{Introduction}
\label{sec:intro}

Systems that can be characterised by a dimensionless parameter $N\gg 1$ 
are of considerable experimental and theoretical significance.
Prominent examples include interacting Bose-Einstein condensates 
and baryons in the quantum theory of SU($N$)-chromodynamics. 
In the case of Bose-Einstein condensates, $N$ simply counts the
bosons that constitute the system. In quantum chromodynamics, 
colour neutrality of baryons implies that $N$ can be identified with the 
number of valence quarks confined inside the baryons. 
The main amenity offered by large-$N$ systems is a natural expansion 
parameter given by $1/N$.  In quantum chromodynamics this 
expansion parameter has a diagrammatic interpretation as planar dominance,
which has been exploited, for instance, in the $1/N$-expansion
of heavy baryons \cite{tHooft, Witten}.

A generic feature of large-$N$ systems is their non--perturbative character.
Even if the elementary interactions between individual constituents are 
consistent with a weak-coupling regime, the large number of constituents 
can lead to strong collective effects experienced by any individual constituent
amidst the others. This suggests a mean-field description, which is well 
understood in the non-relativistic domain, when the Hartree approximation 
can be applied. Large-$N$ systems in the relativistic domain, however, 
are much less understood from a theoretical point of view. 
Attempts to describe these systems based on, for instance, 
the Dyson-Schwinger equations are usually too complicated to allow
for a consistent approximation scheme. 

The purpose of this article is to provide an analytical and quantitative framework for realising 
a mean-field description of large-$N$ systems in the relativistic domain. In the approach presented here, 
the mean field is provided by a non-trivial vacuum structure causing in-medium modifications
of the constituent dynamics that can be related to collective binding effects. 
At this level, the bound-state description is similar to the one developed by 
Shifman, Vainshtein \& Zakharov for using quantum chromodynamics as a predictive theory of hadrons. 
Besides the celebrated quark-hadron duality, certain vacuum condensates (Lorentz- and gauge-invariant 
compositions of fields in the normal-ordering prescription)  
of quarks and gluons \cite{Shifman, Shifman1, QuarkHadron} are central concepts in their approach. 
These condensates parametrise the non-trivial vacuum structure of quantum chromodynamics
and allow to represent hadron properties at sufficiently low energies to account for confinement. 

In contrast, the approach presented here does not intend to model confinement effects. 
Rather, condensates are used as phenomenological bookkeeping devices to parametrise the mean field 
experienced by individual constituents in large-$N$ systems. Our main objective is to construct 
a solid theoretical framework which makes good use of these phenomenological ideas. 
As will be shown in detail,
this leads to a representation of relativistic quantum bound-states qualifying as large-$N$ systems 
in terms of an auxiliary current. Such a representation is valid both for the asymptotic framework
pertinent to the scattering matrix, as well as for the construction of kinematical states associated
with large-$N$ systems. Thus, with the aide of the auxiliary current, the corresponding 
bound states can be reduced in the sense of Nishijima and Lehmann, Symanzik \& Zimmermann \cite{Zimmermann, Nishijima},
as well as in the usual sense of absorption and emission processes.  
Obviously, this is an important prerequisite for calculating static and 
dynamical properties of these bound states.

As an application, 
following a recent proposal put forward in
\cite{Dvali}-\cite{Dvali3}\footnote{For Schwarzschild
black holes in the context of matrix models see \cite{Banks}.},
black holes will be considered as large-$N$ systems at a quantitative 
level strictly following the logic of the general bound-state formalism 
developed in this article. The key idea is to model black holes 
as quantum bound-states of $N\gg 1$ constituents in Minkowski space-time. 
Here, constituents include all graviton polarisations, in particular scalar 
gravitons. In this application, Minkowski space-time is not considered as a
specific background geometry, rather it has the status of a distinguished 
space-time. Of course, Schwarzschild space-times are non-perturbative 
deformations of Minkowski space-time, in the sense that arbitrary many
couplings between gravitons and the associated energy-momentum tensor 
have to be considered \cite{Duff} in order to reproduce this geometry. 
But the bound-state description suggested here goes beyond a purely 
perturbative reconstruction. From a geometrical point 
of view, the condensates represent non-perturbative deformations of 
Minkowski space-time. Furthermore, the description allows to construct
observables sensitive to the constituent structure inside the black hole,
such as the momentum-distribution, the number and 
energy density of the constituents. This leads to a description that is also
complementary to the geometrical picture. It is, however, not equivalent,
since typical quantum corrections are only suppressed by $1/N$,
as opposed to being suppressed exponentially, thus qualifying the notion
of black holes as classical entities. As has been pointed out in \cite{Dvali1},
it is exactly this feature which could shed new light on old problems like the information paradox.

The outline of this article is as follows. In Section 2 we introduce the auxiliary current construction
as a tool of representing bound states in terms of the fields appearing in the microscopic Lagrangian.
In order to give a self-contained discussion, we first explain how these bound states
can be embedded into the asymptotic framework of S-matrix theory as in-states or out-states.
Secondly, we show how the construction can be generalised to situations
where the bound state is not an asymptotic state.
Finally, it is explained how symmetries of the bound state can be implemented
directly in the auxiliary current description.

We proceed by constructing the gauge-invariant constituent distribution functions
of scalars inside the bound state. Although we are ultimately interested in the distribution
of gravitons inside black holes, it suffices to consider scalar distribution at the parton level.
Higher order corrections are, however, sensitive to the constituent polarisations, as will be show. 

In Section 4 we discuss the renormalisation of composite operators at parton level.
We show that in the limit of infinite black hole mass, a consistent renormalisation prescription
at lowest order amounts to setting all loop contributions to zero.
Using this prescription, we calculate observables related to the interior structure of black holes
such as the constituent distribution, energy density and total number of black hole constituents
at the parton level. 
While composite operator renormalisation implies that all loop contributions vanish,
non-triviality of our results suggests that condensation must take place.
These condensates correspond to normal-ordered contributions in Wick's theorem.

In the last Section we discuss how gauge corrections can be taken into account.
In order to highlight the practical value of external field methods in this context,
we calculate a specific diagram, leaving a systematic study of gauge corrections for future work.


\section{Auxiliary current description}\label{acd}
In this section it is shown how gravitational bound states can be described by local
and Lorentz-covariant operators, so-called auxiliary currents. 
The latter are constructed from fields
representing gravitons. 
As a first step, we generalise the asymptotic construction of \cite{Zimmermann, Nishijima}
to gravitational bound-states consisting of a large number of gravitons.
In the second part, we derive a similar representation
for non-asymptotic bound states using auxiliary currents. The resulting representation is therefore not tied
to the scattering matrix theory. Rather, it allows us to define observables connected to the black hole
interior in terms of equal time correlation functions.
Finally, it is explained in detail, how symmetries associated to the bound state
under consideration translate directly into symmetries the auxiliary current has to respect.
\subsection{Asymptotic framework}
Historically, the bound 
state problem was first addressed in the asymptotic framework of quantum field
theory. There it has been shown that microscopic causality allows to describe
asymptotic bound states exactly in the same way as elementary particles.
In other words, the principle of microscopic causality offers no distinction 
(within the scattering-matrix) between elementary and composite particles. 
This is true, in particular, for the reduction formalism given by 
Lehmann, Symanzik and Zimmerman.

In this section, we apply the construction to gravitational bound-states expressed in terms of 
elementary gravitons (including scalar gravitons). 
Notice that since asymptotic states correspond to free particles, both	
elementary as well as composite states should be described by a free wave equation
at spatial and temporal infinity. But then we can define an effective operator for the bound state
which is quantised in a similar way as an elementary field. Thus, at least in the framework of 
the scattering matrix, the reduction formalism should not differ from 
that of a free particle (except for normalisation factors
due to the composite nature of the bound state.)
Having this intuitive picture in mind, let us make these statements more rigorous.

Consider gravitons described by a rank-two Lorentz-tensor field $h(x)$ and demand that 
the principle of microscopic causality holds, and that the spectral condition
is obeyed. For simplicity we assume there is only one spin-zero bound state 
$|\mathcal{B}\rangle$ with mass $M$, and $\langle\Omega|h(x)|\Phi\rangle\neq 0$
if $|\Phi\rangle$ is a momentum eigenstate corresponding to vanishing mass and spin-two,
but $\langle\Omega|h(x)|\mathcal{B}\rangle \equiv 0$ and
$\langle\Omega|h(x_1)\cdots h(x_N)|\mathcal{B}\rangle \neq 0$.
Here, $|\Omega\rangle$ denotes the unique ground state in Minkowski space-time.
Note that the latter condition has a natural interpretation using a Fock space representation of the bound state.
Indeed, if the spectrum contains bound states and single-particle states, then it is always possible
to express the former in terms of the latter. Schematically, we can thus write $| \mathcal{B} \rangle = \sum_i \alpha_i | i \rangle$,
where $| i \rangle$ are Fock basis states and $| \alpha_i |^2$ is the probability to find the bound state $| \mathcal{B} \rangle$
in the Fock state $| i \rangle$. Note that $\alpha_i \neq 0$ for those states which have the 
same quantum numbers as $| \mathcal{B} \rangle$\footnote{It is very important to stress that this argument is exact
and not restricted to the scattering matrix. 
In particular, the same reasoning can be applied to the representation of non-asymptotic bound states
by auxiliary currents, see the next subsection.}.  

In order to illustrate the construction, consider spin-0 positronium in quantum electrodynamics.
The corresponding quantum state is characterised by its spin, mass and total charge. 
A proper Fock state is given by a state
consisting of an electron and a positron with their spins anti-aligned. Note, however, 
that we could include gauge-invariant combinations
of the abelian field strength in the state construction, as well. 
This would not change the quantum numbers of the state. 
Hence, in general there is a plethora of quantum states with non-vanishing overlap with the 
true bound state. These eigenstates simply differ by their normalisation.
In the case of a Schwarzschild black-hole, 
possible quantum states include spin-0 combinations of gravitons
with total energy equal to the mass of the black hole.

In order to describe the gravitational bound state $|\mathcal{B}\rangle$ in terms of gravitons, we introduce 
the multi-local auxiliary current centred around $x$, 
\begin{eqnarray}
	\mathcal{J}(x,\zeta)
	=
	{\rm T} \mathcal{C}^{\mu_1\nu_1\cdots\mu_N\nu_N} 
	h_{\mu_1\nu_1}(x+\zeta_1)\cdots h_{\mu_N\nu_N}(x+\zeta_N)
	\;, \hspace{0.5cm}
	\sum_{a=1}^N \zeta_a =0 \;,
\end{eqnarray}	
where $\zeta\equiv(\zeta_1,\dots,\zeta_N)$ and
$\mathcal{C}$ denotes the coupling tensor. Asymptotic fields are given by
\begin{eqnarray}
	&& h^{\rm asy}_{\mu\nu}(x)
	=
	h_{\mu\nu}(x)+
	\int {\rm d}^4 y \; G_{\mu\nu}^{\; \; \; \lambda\sigma}(x-y) T_{\lambda\sigma}(y)
	\;, \nonumber \\
	&& \mathcal{J}^{\rm asy}(x,\zeta)
	=
	\mathcal{J}(x,\zeta)+
	\int {\rm d}^4 y \; \mathcal{G}(x-y) \mathcal{T}(y,\zeta)
	\; ,
\end{eqnarray}		
with $G,\mathcal{G}$ denoting the retarded and advanced Green functions 	
for the incoming and outgoing fields, respectively, with the source operators
\begin{eqnarray}
	&& T_{\mu\nu}(x) = \mathcal{E}_{\mu\nu}^{\; \; \; \alpha\beta} h_{\alpha\beta}(x)
	\;, \nonumber \\
	&& \mathcal{T}(x,\zeta) = \left(\Box - M^2\right) \mathcal{J}(x,\zeta) = 0
	\; .
\end{eqnarray}	
Here, $\mathcal{E}$ denotes the wave operator for the gravitons. Clearly, 
\begin{eqnarray}
	\label{asyeoms}
	&& \mathcal{E}_{\mu\nu}^{\; \; \; \alpha\beta} h^{\rm asy}_{\alpha\beta}(x) = 0 
	\;, \nonumber\\
	&&  \left(\Box - M^2\right) \mathcal{J}^{\rm asy}(x,\zeta) = 0 
	\; .
\end{eqnarray}	
As a consequence of (\ref{asyeoms}) an the covariance properties of the asymptotic
field operators, their expectation value with respect to the ground state vanishes,
$\langle\Omega|h^{\rm asy}(x)|\Omega\rangle=0$ and 
$\langle\Omega|\mathcal{J}^{\rm asy}(x,\zeta)|\Omega\rangle=0$.  
Furthermore, $h^{\rm asy}$ and $\mathcal{J}^{\rm asy}$ satisfy the usual
asymptotic condtions, for instance 
\begin{eqnarray}
	\label{asycond}
	\lim_{x_0\rightarrow\pm\infty} 
	\int_{\Sigma_{x_0}} {\rm d}^3x\; 
	\mathcal{J}(x,\zeta)\overleftrightarrow{\partial_0} F^*(x)
	=
	\int {\rm d}^3x\; 
	\mathcal{J}^{\rm asy}(x,\zeta)\overleftrightarrow{\partial_0} F^*(x) \; ,
\end{eqnarray}	
for any normalizable solution $F$ of the free Klein-Gordon equation, and $\overleftrightarrow{\partial_0}\equiv \partial_0 - \overleftarrow{\partial_0}$
with $\overleftarrow{\partial_0}$ acting to the left. The label $\Sigma_{x_0}$ 
denotes the spatial hypersurface at time $x_0$ according to an inertial observer,
while  
the right hand side of (\ref{asycond}) is time-independent.

The commutators of the incoming and outgoing fields coincide and are c-numbers. 
We focus on the bound state. It is convenient to expand $\mathcal{J}$ and
$\mathcal{J}^{\rm asy}$ with respect to a complete orthonormal system 
$\{F_{\bf k}(x)\}$ of positive frequency solutions of $(\Box-M^2)F(x)=0$:
\begin{eqnarray}
	&&\mathcal{J}(x,\zeta) 
	=
	\sum_\alpha \; \left(F_{\alpha}(x) \mathcal{J}_{{\alpha}+}(x_0,\zeta)  
	+ F_{\alpha}^{\; *}(x) \mathcal{J}_{{\alpha}-}(x_0,\zeta)
	\right)
	\;, \nonumber \\
	&&\mathcal{J}^{\rm asy}(x,\zeta) 
	=
	\sum_\alpha \; \left(F_{\alpha}(x) 
	\mathcal{J}^{\rm asy}_{{\alpha}+}(x_0,\zeta)  
	+ F_{\alpha}^{\; *}(x) \mathcal{J}^{\rm asy}_{{\alpha}-}(x_0,\zeta)
	\right) \; .
\end{eqnarray}	
The coefficients are given by
\begin{eqnarray}
	\mathcal{J}_{{\alpha}\pm}(x_0,\zeta)
	=
	\mp {\rm i} \int_{\Sigma_{x_0}} {\rm d}^3x \; 
	\mathcal{J}(x,\zeta) \overleftrightarrow{\partial_0} F_{{\alpha}\mp}(x)
	\;, 
\end{eqnarray}	
where $F_{{\alpha}-}\equiv F_{\alpha}$ and $F_{{\alpha}+}\equiv F_{\alpha}^*$,
and similar expressions for $\mathcal{J}^{\rm asy}_{{\alpha}\pm}$.
In order to show that $[\mathcal{J}^{\rm in}(x,\zeta),\mathcal{J}^{\rm in}(y,\eta)]$
and $[\mathcal{J}^{\rm out}(x,\zeta),\mathcal{J}^{\rm out}(y,\eta)]$ coincide, we start
from  
\begin{eqnarray}
	\label{iden}
	&&\int {\rm}d^4 x \; {\rm d} ^4 y\; F^*_{\alpha}(x) F_{\beta}(y)
	(\Box_x-M^2)(\Box_y-M^2) {\rm T}\mathcal{J}(x,\zeta)\mathcal{J}(y,\eta) =
	\nonumber \\
	&&
	\int {\rm}d^4 y \; {\rm d} ^4 x\; F^*_{\alpha}(x) F_{\beta}(y)
	(\Box_x-M^2)(\Box_y-M^2) {\rm T}\mathcal{J}(x,\zeta)\mathcal{J}(y,\eta) \; .
\end{eqnarray}	
Strictly speaking, this is not an identity. However, as a consequence
of causality and the spectral condition, the ground-state expectation 
values $\langle\Omega|h(x_1)\cdots h(x_m) {\rm T}[\mathcal{J}(x,\zeta)\mathcal{J}(y,\eta)]
h(y_1)\cdots h(y_n)|\Omega\rangle$ are boundary values of analytical 
Wightman functions. Thus, it can be shown that interchanging the integrations is
indeed justified. Then, (\ref{iden}) holds between any states, since any 
state can be represented by a superposition of states 
$h(x_1)\cdots h(x_n)|\Omega\rangle$. 

Using Green's theorem, it follows that
\begin{eqnarray}
	&&-{\rm i}\int {\rm d}^4y \; F_{\beta}(y) (\Box_y-M^2)T\mathcal{J}(x,\zeta)\mathcal{J}(y,\eta)
	\nonumber \\
	&&=
	-{\rm i}\int {\rm d}y_0\; \partial_0
	\int {\rm d}^3 y \; 
	{\rm T}\mathcal{J}(x,\zeta)\mathcal{J}(y,\eta)\overleftrightarrow{\partial_0} F_{\beta}(y) \;.
\end{eqnarray}
But this is just $\mathcal{J}(x,\zeta)\mathcal{J}_{{\beta}-}^{\rm in}(\eta)
-\mathcal{J}_{{\beta}-}^{\rm out}(\eta)\mathcal{J}(x,\zeta)$. Proceeding in the same way with 
the x-integration, we find for the left hand side of (\ref{iden})
\begin{eqnarray}
	&&\int {\rm}d^4 x \; {\rm d} ^4 y\; F^*_{\alpha}(x) F_{\beta}(y)
	(\Box_x-M^2)(\Box_y-M^2) {\rm T}\mathcal{J}(x)\mathcal{J}(y) = 
	\\
	&&
	\mathcal{J}_{{\beta}-}^{\rm out}(\eta)\mathcal{J}_{{\alpha}+}^{\rm out}(\zeta)
	-
	\mathcal{J}_{{\beta}-}^{\rm out}(\eta)\mathcal{J}_{{\alpha}+}^{\rm in}(\zeta)
	-
	\mathcal{J}_{{\alpha}+}^{\rm out}(\zeta)\mathcal{J}_{{\beta}-}^{\rm in}(\eta)
	+
	\mathcal{J}_{{\alpha}+}^{\rm in}(\zeta)\mathcal{J}_{{\beta}-}^{\rm in}(\eta)
	\; . \nonumber
\end{eqnarray}	
For the right hand side, we find a similar expression. As a result,
(\ref{iden}) implies
\begin{eqnarray}
	\left[\mathcal{J}^{\rm in}(x,\zeta),\mathcal{J}^{\rm in}(y,\eta)\right]
	=
	[\mathcal{J}^{\rm out}(x,\zeta),\mathcal{J}^{\rm out}(y,\eta)] \;.
\end{eqnarray}	
The statement that the commutators of the asymptotic fields are c-numbers
can be derived from 
\begin{eqnarray}
	&&\int {\rm}d^4 x \; {\rm d} ^4 y\; F^*_{\alpha}(x) F_{\beta}(y)
	(\Box_x-M^2)(\Box_y-M^2) {\rm T}\mathcal{J}(x,\zeta)\mathcal{J}(y,\eta) h(z) =
	\nonumber \\
	&&
	\int {\rm}d^4 y \; {\rm d} ^4 x\; F^*_{\alpha}(x) F_{\beta}(y)
	(\Box_x-M^2)(\Box_y-M^2) {\rm T}\mathcal{J}(x,\zeta)\mathcal{J}(y,\eta) h(z) \; .
\end{eqnarray}	
Using again Green's theorem, it follows that 
\begin{eqnarray}
	\left[\left[\mathcal{J}_{{\alpha}+}^{\rm in}(\zeta),\mathcal{J}_{{\beta}-}^{\rm in}(\eta)\right],h(z)\right]
	= 0 \;.
\end{eqnarray}	
Therefore, the commutator $[\mathcal{J}^{\rm in}(x,\zeta),\mathcal{J}^{\rm in}(y,\eta) ]$ is a c-number.

In order to determine the commutator, we calculate its expectation values with respect to the ground state.
For this purpose, consider first the matrix elements of $\mathcal{J}^{\rm asy}$ between the ground state
and an arbitrary state. Let $|{\bf k}\rangle$ be a momentum eigenstate with eigenvalue components $k_\mu$   
and rest mass given by $k^2=-M^2$. Then, 
\begin{eqnarray}
	\langle\Omega| \mathcal{J}(x,\zeta) |{\bf k}\rangle 
	= (2\pi)^{-3/2} \mathcal{N}_{\bf k}(\zeta) \exp{({\rm i}k\cdot x)}
	\;,
\end{eqnarray}	
by translation invariance. Note that the amplitude $\mathcal{N}_{\bf k}(\zeta)$ 
depends on the relative coordinates 
$\zeta_1,\dots,\zeta_N$  with respect to the centre $x$, 
\begin{eqnarray}
	&& \mathcal{N}_{\bf k}(\zeta)
	=
	(2\pi)^{3/2} \langle\Omega| 
	{\rm T}\mathcal{C}^{\mu_1\nu_1\cdots\mu_N\nu_N} h_{\mu_1\nu_1}(\zeta_1)\cdots  
	h_{\mu_N\nu_N}(\zeta_N)|{\bf k}\rangle \;, 
\end{eqnarray}	
and the sum over all relative coordinates vanishes by definition. Clearly,
\begin{eqnarray}
	\left(\Box_x-M^2\right) \langle\Omega|\mathcal{J}(x,\zeta)|{\bf k}\rangle = 0 \;.
\end{eqnarray}	
From this and the definition of the asymptotic bound state it follows that
$\langle\Omega|\mathcal{J}^{\rm asy}(x,\zeta)|{\bf k}\rangle = \langle\Omega|\mathcal{J}(x,\zeta)|{\bf k}\rangle$.
On the other hand, if $|{\bf k}\rangle$ is a momentum eigenstate with eigenvalue components $k_\mu$ 
but rest mass $k^2\neq -M^2$, then
$(k^2+M^2)\langle\Omega|\mathcal{J}^{\rm asy}(x,\zeta)|{\bf k}\rangle$ $=$ 
$-(\Box-M^2) \langle\Omega|\mathcal{J}^{\rm asy}(x,\zeta)|{\bf k}\rangle$ $=0$. As a consequence,
$\langle\Omega|\mathcal{J}^{\rm asy}(x,\zeta)|{\bf k}\rangle =0$ in this case. Therefore, 
\begin{eqnarray}
	\langle\Omega|\mathcal{J}^{\rm asy}(x,\zeta)\mathcal{J}^{\rm asy}(y,\eta)|\Omega\rangle
	&=& \int \tfrac{{\rm d}^3k}{2k_0} \;
	\langle\Omega|\mathcal{J}^{\rm asy}(x,\zeta)|{\bf k}\rangle\langle{\bf k}|
	\mathcal{J}^{\rm asy}(y,\eta)|\Omega\rangle
	\nonumber \\
	&=&
	 \int \tfrac{{\rm d}^3k}{2k_0} \; \exp{({\rm i k\cdot (x-y)})} 
	\mathcal{N}_{\bf k}(\zeta) \mathcal{N}_{\bf k}(\eta) \; .
\end{eqnarray}	
Introducing the Fourier-transform $\mathcal{J}^{\rm asy}(k,\zeta)$  of the asymptotic auxiliary currents by
\begin{eqnarray}
	\mathcal{J}^{\rm asy}(x,\zeta)
	=
	\int \tfrac{{\rm d}^4k}{(2\pi)^{3/2}} \; \exp{({\rm i}k\cdot x)} \delta(k^2+M^2) 
	\mathcal{J}^{\rm asy}(k,\zeta) \;,
\end{eqnarray}	 
as well as absorption and emission operators, 
$\mathcal{J}^{\rm asy}_\pm({\bf k},\zeta)\equiv \mathcal{J}^{\rm asy}(\pm k,\zeta)$ for
$k_0=\pm\sqrt{{\bf k}^2+M^2}$, we find the usual commutation relations,
\begin{eqnarray}
	\label{commasy}
	\left[\mathcal{J}_+^{\rm asy}({\bf k},\zeta),\mathcal{J}^{\rm asy}_-({\bf q},\eta)\right]
	=
	2 k_0({\bf k}) \; \mathcal{N}_{\bf k}(\zeta) \mathcal{N}_{\bf q}(\eta) \; 
	\delta^{(3)}({\bf k}-{\bf q})
	\;, 
\end{eqnarray}	
and all other commutators vanish. Similarly, we can show that commutators between 
emission/absorption operators of elementary fields and auxiliary currents vanish. 
Furthermore, a local version of (\ref{commasy}) is readily derived. Introduce 
\begin{eqnarray}
	\label{limJ}
	\mathcal{J}^{\rm asy}(x)
	:=
	\lim_{\zeta\rightarrow 0}
	\mathcal{N}_{\bf 0}^{\; -1}(\zeta) \mathcal{J}^{\rm asy}(x,\zeta)
	\; ,
\end{eqnarray}	
where $\mathcal{N}_{\bf 0}(\zeta)
\equiv (2\pi)^{3/2}\langle\Omega|\mathcal{J}^{\rm asy}(0,\zeta)|{\bf 0}\rangle$, 
and $|{\bf 0}\rangle$ denotes the bound state at rest. 
The limiting process in the definition (\ref{limJ}) makes sense since the ratio
$\mathcal{J}^{\rm asy}_{\pm}({\bf k},\zeta)/\mathcal{N}_{\bf k}(\zeta)$
can be shown to be $\zeta$-independent and 
$\mathcal{N}_{\bf k}(\zeta)$ becomes ${\bf k}$-independent in the limit $\zeta\rightarrow 0$.
Hence, in the limit $\zeta,\eta\rightarrow 0$ the local version of
the commutator (\ref{commasy}) is given by
\begin{eqnarray}
	\left[\mathcal{J}_+^{\rm asy}({\bf k}),\mathcal{J}^{\rm asy}_-({\bf q})\right]
	=
	2 k_0({\bf k}) \;
	\delta^{(3)}({\bf k}-{\bf q}) \; .
\end{eqnarray}
As a consequence, the commutator of two local asymptotic auxiliary currents is given by
the usual Pauli-Jordan function, and local asymptotic auxiliary currents satisfy 
the free equation of motion. If $|{\bf k}\rangle$ denotes a gravitational bound state
with four momentum $k$ on-shell, $k^2=-M^2$, then 
$\langle\Omega|\mathcal{J}^{\rm asy}(x)|{\bf k}\rangle=(2\pi)^{-3/2}\exp{({\rm i}k\cdot x)}$.
The local asymptotic currents $\mathcal{J}^{\rm asy}(x)$ are given in terms
of elementary graviton fields by
\begin{eqnarray}
	\mathcal{J}^{\rm asy}(x)
	&=& 
	\lim_{\zeta\rightarrow 0}\mathcal{N}_{\bf 0}^{\; -1}(\zeta)\Bigg\{
	\int {\rm d}^4y \; \mathcal{G}(x-y) \mathcal{T}(y,\zeta)+\nonumber \\
	&&\hspace{1cm}{\rm T}\; 
	\mathcal{C}^{\mu_1\nu_1\dots\mu_N\nu_N} h_{\mu_1\nu_1}(x+\zeta_1)\cdots
	h_{\mu_N\nu_N}(x+\zeta_N)\Bigg\} \; ,\\
	\mathcal{T}(x,\zeta)
	&=&
	\left(\Box-M^2\right){\rm T}\; 
	\mathcal{C}^{\mu_1\nu_1\dots\mu_N\nu_N} h_{\mu_1\nu_1}(x+\zeta_1)\cdots
	h_{\mu_N\nu_N}(x+\zeta_N) \; .\nonumber
\end{eqnarray}	

A complete orthonormal system for the whole Hilbert space can be constructed 
from the emission operators corresponding to elementary gravitons and the 
emission operators $\mathcal{J}^{\rm asy}_{-}({\bf k})$ for gravitational 
bound states of momentum $k$. For instance, an in state vector corresponding  
to a single bound state of momentum $k$ is given by
$|k \; {\rm in}\rangle = \mathcal{J}^{\rm in}_-({\bf k})|\Omega\rangle$.
The asymptotic condition (\ref{asycond}) for plane waves reads
\begin{eqnarray}
		\label{wasyc}
		\lim_{x_0\rightarrow\pm\infty}
		\int_{\Sigma_{x_0}}{\rm d^3}x \; 
		\mathcal{J}(x,\zeta) \overleftrightarrow{\partial_0}
		{\rm e}^{\mp{\rm i}k\cdot x}
		=
		\int {\rm d^3}x \; \mathcal{J}^{\rm asy}(x,\zeta) 
		\overleftrightarrow{\partial_0}{\rm e}^{\mp{\rm i}k\cdot x} \;.
\end{eqnarray}	
As usual, modulo a disconnected contribution (when evaluated in states), this gives
\begin{eqnarray}
	\mathcal{J}^{\rm in}_-({\bf k},\zeta)
	&=&
	-{\rm i}\int {\rm d}^4 x\; 
	\partial_0\left(\mathcal{J}(x,\zeta) \overleftrightarrow{\partial_0}
	\tfrac{{\rm e}^{{\rm i}k\cdot x}}{(2\pi)^{3/2}}
	\right)\; ,
\end{eqnarray}
and after two integrations by parts, we find the reduction formula relating
an asymptotic gravitational bound state with the ground state,  
\begin{eqnarray}	
	|k \; {\rm in}\rangle 
	&=&
	\tfrac{{\rm i}}{(2\pi)^{3/2}\langle\Omega|\mathcal{J}(0,0)|{\bf 0}\rangle}
	\int {\rm d}^4 x\; \tfrac{{\rm e}^{{\rm i}k\cdot x}}{(2\pi)^{3/2}}
	\left(\Box - M^2\right) \mathcal{J}(x,0) |\Omega\rangle \; .
\end{eqnarray}	

For the asymptotic framework pertinent to the scattering matrix we are only interested 
in the centre of mass coordinates of the auxiliary currents. It suffices to construct 
local field operators representing the bound states by taking the limit $\zeta\rightarrow 0$
of  the multi-local auxiliary current $\mathcal{J}(x,\zeta)$. We assume the existence of
\begin{eqnarray}
	\label{locaux}
	\mathcal{J}(x)
	\equiv
	\lim_{\zeta\rightarrow 0}
	\frac{\mathcal{J}(x,\zeta)-\langle\Omega|\mathcal{J}(0,\zeta)|\Omega\rangle}
		{(2\pi)^{3/2}\langle\Omega|\mathcal{J}(0,\zeta)|{\bf 0}\rangle}
	\; ,
\end{eqnarray}	
where $|\bf 0\rangle$ denotes the bound state at rest.
The local auxiliary current $\mathcal{J}(x)$ transforms covariant with respect to the
inhomogeneous Lorentz group, and $[\mathcal{J}(x),\mathcal{J}(y)]=0$ for
$\|x-y\|^2>0$. Furthermore, it satisfies the asymptotic conditions, i.e.
\begin{eqnarray}
	\lim_{x_0\rightarrow\pm\infty}
	\int_{\Sigma_{x_0}}{\rm d^3}x \; \mathcal{J}(x) \overleftrightarrow{\partial_0}F^*(x)
	=
	\int {\rm d^3}x \; \mathcal{J}^{\rm asy}(x) \overleftrightarrow{\partial_0}F^*(x) \;,
\end{eqnarray}		
for normalisable solutions $F$ of the free Klein-Gordon equation. Note that the right hand
side is time independent. 

In summary, the local auxiliary current $\mathcal{J}(x)$ transforms covariant under the
inhomogeneous Lorentz transformation, respects causality and satisfies the asymptotic
conditions exactly in the same way as the local fields representing elementary particles. 
Hence, the Lehmann-Symanzik-Zimmermann reduction formalism can be used 
to get the usual expansion of the scattering matrix:
\begin{eqnarray}
	S&=&\sum_{m,n\in\mathbb{N}} \tfrac{(-{\rm i})^{m+n}}{m!n!}
	\int {\rm d}^4 x_1\cdots {\rm d}^4 x_m \int {\rm d}^4 y_1\cdots {\rm d}^4 y_n\; 
	\Box_{x_1}\cdots \Box_{x_m}	\nonumber\\
	&& \langle\Omega|{\rm T}h(x_1)\cdots h(x_m) \mathcal{J}(y_1)\cdots\mathcal{J}(y_n) |\Omega\rangle
	\left(\overleftarrow{\Box}_{y_1}-M^2\right)\cdots \left(\overleftarrow{\Box}_{y_n}-M^2\right) \nonumber\\
	&& {\bf :}h^{\rm in}(x_1)\cdots h^{\rm in}(x_m) \mathcal{J}^{\rm in}(y_1)\cdots\mathcal{J}^{\rm in}(y_n){\bf :}
	\; .
\end{eqnarray}		

There is one important difference between fields representing elementary particles and 
auxiliary currents representing bound states. If $\mathcal{J}(x)$ is the local auxiliary current corresponding to 
a bound state composed of $N$ gravitons described by rank-2 Lorentz tensors $h$, it is possible to represent
$\mathcal{J}(x)$ as a monom in $h$, 
\begin{eqnarray}
	\label{rela}
	\mathcal{J}(x)
	=
	\mathcal{N}^{-1/2}\left(\mathcal{J}(x,0)-\mathcal{V}\right)
	\; ,
\end{eqnarray}		
with the renormalisation constants 
\begin{eqnarray}
	\mathcal{N}
	&=&
	-{\rm i} \int {\rm d}x_0\; \exp{(-{\rm i}Mx_0)} \langle\Omega| 
	{\rm T} \mathcal{J}(0,0)\mathcal{J}(x,0)|\Omega\rangle
	\nonumber \\
	\mathcal{V}
	&=&
	\langle\Omega|\mathcal{J}(0,0)|\Omega\rangle
	\; .
\end{eqnarray}	
Equation (\ref{rela}) may be imposed as an additional condition beyond the principles fields representing
elementary particles have to satisfy (covariance, causality and asymptotic conditions). 

\subsection{General reduction}
The Fock space representation presented above 
allows to relate observables that characterise bound states of graviton constituents 
to scattering processes involving these bound states. 
The representation of bound states by multi-local auxiliary currents, however,
is not restricted to the asymptotic framework pertinent to scattering theory. In fact, the idea
of representing bound state properties by perturbative degrees of freedom (local bookkeeping
devices) is completely generic, as long as it suffices to consider bound states at the 
purely kinematical level. 
This fact is again related to the Fock-space expansion of bound states as discussed above.

Let $|k\rangle$ denote a bound state with four-momentum $k, k^2=-M^2$,
and an unspecified list of quantum numbers compatible with that of the bound state. This state can be related to the 
ground state as follows:
\begin{eqnarray}
	|k\rangle
	&=&
	 2k_0 \Gamma^{-1}\int {\rm d}^3x \; \tfrac{{\rm e}^{{\rm i}k\cdot x}}{(2\pi)^{3/2}}\; 
	\mathcal{J}(x) |\Omega\rangle
	\;, 
\end{eqnarray}	
where $\Gamma\equiv (2\pi)^{3/2}\langle\Omega|\mathcal{J}(0,0)|{\bf 0}\rangle$, and 
$\mathcal{J}(x)$ denotes the local auxiliary current as given in (\ref{locaux}) with 
the current normalisation $\Gamma$ factored out. 
A generic gravitational bound state $|\mathcal{B}\rangle$ is a superposition of momentum eigenstates
with appropriate mass and quantum numbers,
\begin{eqnarray}
	|\mathcal{B}\rangle
	&=&
	\Gamma^{-1}\int {\rm d}^3k \; \mathcal{B}(k)
	\int {\rm d}^4x \; \tfrac{{\rm e}^{{\rm i}k\cdot x}}{(2\pi)^{3/2}}\; \mathcal{J}(x)|\Omega\rangle \; .
\end{eqnarray}	
Here, $\mathcal{B}(k)$ is the wave function of the bound state in momentum space.

For later convenience, let us already mention that
requiring $\langle\mathcal{B}|\mathcal{B}\rangle=1$, gives the current normalisation $\Gamma$ as
\begin{eqnarray}
	\label{Gamma2}
	\Gamma^2
	&=&
	\left(\tfrac{N}{M}\right)^2 \langle\Omega|\hspace{-0.1cm}:\hspace{-0.1cm}h^{2(N-1)}(0)\hspace{-0.1cm}:
	\hspace{-0.1cm}|\Omega\rangle
	\int {\rm d}^3p \; \left|\mathcal{B}(p)\right|^2 
	\; \delta_\mathcal{B} \; ,
\end{eqnarray}	
with $\delta_\mathcal{B}$ indicating that $|\mathcal{B}\rangle$ and its dual
are localised on the same spatial hypersurface.
This result will be derived in Section 4. As will be shown there, (\ref{Gamma2})  follows in the limit 
$M/\mu\rightarrow\infty$, where $\mu$ denotes an arbitrary energy scale, 
which can only be considered together with $N\rightarrow\infty$ such that $M/\mu/N$ becomes 
constant.


\subsection{Isometries and symmetries of auxiliary currents}\label{Symm}	
Solutions of general relativity are usually classified with respect to their isometries. 
This raises the question, how space-time isometries can be implemented in the auxiliary current 
description of the corresponding quantum bound-state  $|\mathcal{B} \rangle$.
An obvious requirement is that $|\mathcal{B} \rangle$ should be left invariant under 
the action of the isometry generators. At the same time, since the bound-state breaks some of the isometries 
characterising Minkowski space-time explicitly, it should transform non-trivially 
under the generators of the broken isometries. 
Let $\mathcal{G}$ denote a collection of unbroken generators 
and $\mathcal{H}$ a collection of broken generators. Then, 
\begin{eqnarray}
\label{transformation}
&&\mathcal{G} | \mathcal{B} \rangle = | \mathcal{B} \rangle \;, \hspace{0.5cm}
\mathcal{H} | \mathcal{B} \rangle = | \mathcal{B}' \rangle \; ,
\end{eqnarray}
with $| \mathcal{B} \rangle \neq | \mathcal{B}' \rangle$.
In the following, we investigate how these transformation properties 
are realised at the level of the auxiliary currents associated with the 
quantum bound-states. This requires to consider the action of the generators 
on $\mathcal{J}(x)|\Omega\rangle$ at every space-time location.
Since the ground state is left invariant by all generators of the Poincare group,
the transformation properties of the bound states are captured in the space-time
dependence of the auxiliary currents.   
In particular, denoting infinitesimal transformations by $\mathcal{G} \simeq 1 + \delta_{\mathcal{G}}$
and $\mathcal{H} \simeq 1 + \delta_{\mathcal{H}}$  respectively, we have
\begin{eqnarray}
\label{infinitesimal}
&&\delta_{\mathcal{G}} \mathcal{J}(x) = [\mathcal{G}, \mathcal{J}(x)] = 0 \; ,\hspace{0.5cm}
\delta_{\mathcal{H}} \mathcal{J}(x) = [\mathcal{H}, \mathcal{J}(x)] \neq 0 \; .
\end{eqnarray}
The right-hand side of (\ref{infinitesimal}) can be translated into a differential equation
determining the space-time dependence of the auxiliary currents for a given background.
In this way the classical background isometries can be implemented in the auxiliary
current description of the corresponding quantum bound-state.

As an example, consider spherical symmetric space-times and, in particular, Schwarzschild
black-holes. When considered as bound states, these solutions are clearly invariant 
under spatial rotations and time-translations. 
The corresponding generators can be represented as
$\mathcal{G}_{ab} = x_a \partial_b - x_b \partial_a$ and $\mathcal{G}_t = \partial_t$, respectively.
Here, $a, b\in\{1,2,3\}$ index spatial coordinates and $t$ is the Minkowski time-coordinate
of an inertial observer.
From (\ref{infinitesimal}) it follows that the auxiliary current representing a Schwarzschild black-hole 
can only depend on the spatial distance $|{\bf{ r}}|$ from the origin. In consideration of the 
auxiliary current construction, this dependence descents to the individual field operators in the composition.

While such symmetry restrictions are to be expected, it is desirable to perform all calculations 
in a manifest Lorentz covariant framework, and to reduce to the actual isometries only at the end. 
Fortunately, we can proceed in exactly such a way
by virtue of Ward's identity.
Using the invariance of the state $| \mathcal{B} \rangle$ under the unbroken generators, 
Ward's identity leads to 
\begin{eqnarray}
      0=\langle\mathcal{B}|\partial_\mu j^\mu | \mathcal{B}\rangle
      =\langle\mathcal{B}|\delta_{\mathcal{G}} \mathcal{O} | \mathcal{B}\rangle
      =\delta_{\mathcal{G}} \langle \mathcal{B}|\mathcal{O} | \mathcal{B}\rangle \; .
      \label{ward}
\end{eqnarray}
Here, $j$ denotes the conserved current associated with the isometries 
(not to be confused with the auxiliary current $\mathcal{J}$).
In practice (\ref{ward}) implies that observables can be calculated
in a fully Lorentz covariant way and the symmetry constraints 
can be imposed at the end of the calculation.


\section{Constituent distribution function}
In this section we review the construction of a gauge-invariant
operator that measures the constituent distribution in a given 
bound state $|\mathcal{B}\rangle$. The construction can be viewed as the analogue of the gauge-invariant
completion of the quark distributions in the context of quantum chromodynamics \cite{Collins}. 
For simplicity, the construction presented here
will be restricted to distributions of real massless scalars, $h(x)$, in the presence of gravity.
As will be shown in the next section, at the parton level, the only difference
between a graviton distribution and a scalar distribution is a numerical pre-factor.
Thus, when working at the parton level, it suffices to consider scalar distributions.
Note, however, that the difference becomes important beyond the parton level. 
Then, the scalars are coupled to gravity and the scalar as well as the graviton distribution
can be investigated separately. A physical situation where this can arise is the collaps
of a spherical shell consisting of scalar constituents.
The resulting system will contain scalars as well as longitudinal gravitons. 

Introducing the Fourier-transform $h(k)$ of the free constituent field by
\begin{eqnarray}
	h(x)
	&=&
	\int {\rm d}^4 q \;\tfrac{{\rm e}^{{\rm i}q\cdot x}}{(2\pi)^{3/2}} \delta(q^2) h(q) \; ,
\end{eqnarray}	
as well as absorption and emission operators, $a({\bf q})\equiv h(q)$ and $a^\dagger({\bf q})\equiv h(-q)$
for $q_0=\pm |{\bf q}|$, we have the usual commutator relations
$[a({\bf q}), a^\dagger({\bf k})]=2q_0({\bf q})\delta^{(3)}({\bf q}-{\bf k})$, and all other commutators
vanish. Explicitly,  
\begin{eqnarray}
	a^\dagger({\bf q}) &=& {\rm i}\int_\Sigma {\rm d}^3 x\; \tfrac{{\rm e}^{{\rm i}q\cdot x}}{(2\pi)^{3/2}}
	 \;  \overleftrightarrow{\partial_0} \; h(x)
	  \; ,
\end{eqnarray}	
whereby the on-shell condition is implied. 

The occupation number density in a momentum-space 
volume ${\rm d}^3q$ centred around ${\bf q}$ is  $n({\bf q})\equiv {\rm d}\mathcal{N}_{\rm c}/{\rm d}^3 q$,
where $\mathcal{N}_{\rm c}$ denotes the total constituent number, can be expressed as 
\begin{eqnarray}
	\label{numb}
	n({\bf q})
	&=&
	\left( 2|{\bf q}|\right)^2 \int {\rm d}^3 x {\rm d}^3 y\; 
	\tfrac{{\rm e}^{{\rm i}q\cdot (x-y)}}{(2\pi)^3} \; h(x) h(y)
	\; .
\end{eqnarray}
Notie that while $N$ denotes the number of fields composing the auxiliary current, $\mathcal{N}_{\rm c}$
counts the total number of constituents (including virtual ones). As will be explained in detail below,
the effect of virtuality in our approach is accounted for in terms of vacuum condensates even at the parton level.
Thus, in general, $N$ and $\mathcal{N}_{\rm c}$ do not coincide.
Let us stress again that a similar statement can be made in quantum chromodynamics. 
Indeed, in order to describe
hadrons, in principle auxiliary currents constructed solely from valence quarks can be used.
Nevertheless, the gauge-invariant distribution of gluons in the hadron can be calculated, as well. 
The reason is that due to interactions, virtual gluons are sourced.
Thus, integrating the distribution over all momenta, a non-vanishing
total number of gluons inside the hadron can be defined.

Let us come back to equation (\ref{numb}).
As an example, consider momentum eigenstates $|p\rangle$ with four momenta $p, p^2=-m^2$,
and a state $|\Psi\rangle$ corresponding to a single free elementary particle of mass $m$ (with 
the case $m=0$ included).  Denoting by
$\Psi(p)\equiv \langle p|\Psi\rangle$, we find for the expectation of the number density in the state $|\Psi\rangle$,
$\langle \Psi|n({\bf q})|\Psi\rangle\propto |\Psi({\bf q})|^2$. Intuitively, this relates the number density to the field intensity. 

Next consider a bound state $|\mathcal{B}\rangle$, corresponding to a composite object
of mass $M$ with wave function $\mathcal{B}(p)$, where $p$ denotes its four-momentum. 
Let us introduce $r\equiv (x-y)/2$ and $R\equiv (x+y)/2$ in (\ref{numb}) and perform similarity transformations
using the appropriate unitary representations of space-time-translation, 
$h(x)h(y)=U^{-1}(R-r)h(r)h(0)U(R-r)$. If evaluated in the state $|\mathcal{B}\rangle$, the ${\bf R}$-integration
becomes trivial and allows to eliminate the dependence on $R_0$. We find,
\begin{eqnarray}
	\langle\mathcal{B}|n({\bf q})|\mathcal{B}\rangle
	=
	\left( 2|{\bf q}|\right)^2
	\int \tfrac{{\rm d}^3p}{(2p_0)^2}\; 
	\left|\mathcal{B}({\bf p})\right|^2 \int_\Sigma {\rm d}^3 r \;
	{\rm e}^{{\rm i}q\cdot r} \; \langle p|h(r)h(0)|p\rangle
	\; .
\end{eqnarray}	
From this we infer that the bi-local operator $\mathcal{O}(r,0)\equiv h(r)h(0)$ 
is the observable in the free theory 
that allows to measure the constituent distribution in the following sense:
\begin{eqnarray}
	 \mathcal{D}(r)
	 &\equiv&
	 \int \tfrac{{\rm d}^3q}{(2\pi)^{3}(2|{\bf q}|)^2} \; {\rm e}^{-{\rm i}q\cdot r} \; 
	  \left\langle \mathcal{B}| n({\bf q}) |\mathcal{B}\right\rangle
	  \nonumber \\
	  &=&
	  \int \tfrac{{\rm d}^3p}{(2p_0)^2}\; \left|\mathcal{B}({\bf p})\right|^2
	  \langle p|h(r)h(0)|p\rangle 
	 \; .
\end{eqnarray}	
Generically, $\mathcal{O}(r,0)$ does not give rise to an observable. This is the case, in particular, 
for gauge theories. The bi-local character of $\mathcal{O}(r,0)$ requires a gauge-invariant 
completion. Such a completion can be constructed by connecting $r$ and the origin with a Wilson line,
i.e.~with a path-ordered exponential of the gauge field. For gravitational interactions of the 
constituent fields $h$, the gauge field $\mathcal{G}$ in question is given by the affine connection $\Gamma$,
in components $\mathcal{G}_\mu\equiv \Gamma^\lambda_{\lambda\mu}$.
Suppose $h$ is minimally coupled to gravity. For convenience, we consider 
$\mathcal{O}(r;y/2)\equiv \mathcal{O}(y+r/2,y-r/2)$.
Then, treating $\mathcal{G}$ as an external field,
the equation of motion for $\mathcal{O}(y;r/2)$ is given by
\begin{eqnarray}
	\label{eom}
	\left(-\Box + \mathcal{G} \cdot\partial\right) \mathcal{O}(y;r/2) 
	= \delta^{(4)}(r).
\end{eqnarray}	
This equation can be solved by iteration. A detailed derivation can be found in \ref{Wilson}.
The result is 
\begin{eqnarray}
	\label{O}
	\mathcal{O}(y;r/2)
	=
	\mathcal{P}\exp{\left(-\int\hspace{-0.35cm}C {\rm d}z^\lambda \mathcal{G}_\lambda(z)\right)}
	\; \mathcal{O}^{(0)}(y;r/2),
\end{eqnarray}	
where $C$ denotes the contour given by the path $z:[0,1]\rightarrow \mathbb{R}^4\;,
u\rightarrow z(u):=y-(1-2 u)r/2$, $\mathcal{P}$ refers to path ordering
along this contour, and $\mathcal{O}^{(0)}(y;r/2)={\rm T}h(y+r/2)h(y-r/2)$. 
Note that (\ref{O}) is an exact statement on the light-cone.

\section{Composite operator renormalisation at parton-level}\label{cor}
In section \ref{acd} we showed that black-holes can be descried by local auxiliary currents 
composed of graviton fields in the asymptotic framework pertinent to scattering theory. 
Consider, in particular, the scattering of a probe particle on a black-hole given 
as a quantum bound-state described by such a local auxiliary current. 
The associated cross section factorizes in a term (Wilson coefficient) that can be calculated 
using standard perturbation theory, and the distribution function 
$\langle \mathcal{B} |{\rm{T}} h(x_1) [x_1,x_2] h(x_2) | \mathcal{B} \rangle$  
of the constituent gravitons inside the bound state. Here, $[x_1,x_2]$ 
denotes the Wilson line between the gravitons located at $x_{1,2}$. Since we are mostly interested in
the partonic level in this article, we we can set the path-ordered exponential to the identity.
Clearly, the distribution function carries non-perturbative information, albeit the individual 
interaction between gravitons can be considered weak. Their binding to the bound state is not due to a strong
coupling regime, but rather due to the collective potential an individual graviton experiences in the presence
of all the others. This collective effect can be taken into account in terms of non-vanishing condensates of gravitons
(with respect to the ground state),
which are even present at the parton-level. 
Using the local auxiliary current description, calculating the distribution function
requires, among other things, to calculate the following four-point correlation function:
$\langle \Omega | \mathcal{J}(x) h(x_1) h(x_2) \mathcal{J}(y) | \Omega \rangle.$
In the case of a black hole, $\mathcal{J}$ is a local monomial of graviton fields.
Hence, in order for this correlation function to be meaningful, a renormalisation procedure is required.

In fact, any observable represented by an operator $\mathcal{O}(x_1,\ldots, x_k)$ requires 
renormalisation when evaluated in a bound state which is described by an auxiliary current 
composed of $N>k$ fields.  Let us discuss the renormalisation of composite operators in free field theory,
before turning to actual calculations in the next section. 
Consider a local, operator valued, non-linear functional $\mathcal{F}(x)$ of the field $h(x)$.
This is a slight generalisation of the auxiliary currents we are concerned with and allows, in particular,
to include derivatives of fields. Let $F(x)$ be a local composition of $N$ fields $h$. 
In order to give a regularised expression for these compositions, it suffices \cite{WittenLec}
to properly define the 
ground-state expectation values 
$\langle \Omega |{\rm{T}} h(y_1) \cdots h(y_s) \mathcal{F}(x) | \Omega \rangle$.
These $(s+N)$-point correlation functions can naively be calculated by Wick expansion,
which can be interpreted based on the standard Feynman diagrammatic rules.
Due to the local nature of the composite operator $F$,  the expansion will generate 
self-loops at the location $x$ for $N>s$, each of which leads to the usual divergence. 
In the context of free theory (which we are considering when calculating observables at the parton-level),
there is a straightforward solution to this problem: $F(x)\rightarrow \normord{F(x)}$, where $\normord{}$ denotes 
normal-ordering. Clearly, this removes all self-loops and leads to well-defined expressions for 
all correlation functions.
The normal-ordering prescription exactly corresponds to the regularisation of composite operators 
at the level of a free field theory, with a renormalisation scheme chosen such that any ambiguous 
finite part is set to zero.

For the applications considered in this article, 
we will be interested not only in correlation functions involving one composite operator,
but rather in expressions of the form $\langle \Omega |{\rm{T}}  \mathcal{F}(x) \mathcal{O}(x_1, \ldots, x_k) \mathcal{F}(y) | \Omega \rangle$ with $\mathcal{O}$ denoting an observable constructed
from graviton field operators and their derivatives.
In order to remove the divergencies originating from closed 
loops at $x$ and $y$ we should first of all normal-order 
both composite operators separately. The Wick expansion will now generate loops connecting $x$ and $y$.
The divergencies of these loops can be regulated using standard methods such as dimensional regularisation.
In our case, the auxiliary currents describe the quantum bound-state corresponding to a black hole of mass $M$.
We will be concerned with the limit $M/\mu\rightarrow\infty$, where $\mu$ is any other energy scale.
In this limit, the internal lines corresponding to intermediate propagators 
connecting $x$ and $y$ shrink to a point. As a consequence, the self-loops
are the only contributions that require regularisation, which we demonstrate explicitly in the next section. 
In this situation, we use the following prescription:
$\langle \Omega |{\rm{T}}  \normord{\mathcal{F}(x) \mathcal{O}(x_1, \ldots, x_k) \mathcal{F}(y)} | \Omega \rangle$. 
Wick expanding this $(k+2N)$-correlation function captures the correct physics 
in the limit of arbitrary heavy black holes. 

An evident objection to this type of regularisation is triviality, i.e.~the regularised correlation function 
should vanish for $2N>k$. A purely perturbative calculation of the correlation functions would always
give zero in this case. Hence, within the perturbative framework all observables would be zero for $2N>k$,
which, of course, does not make sense even at the level of free constituents. 
For instance, in SU(N)-quantum 
chromodynamics observables characterising the structure of a bound state can be calculated
at a resolution scale deeply inside 
the regime of asymptotic freedom. Still, a non-trivial description of the bound state
at this resolution scale can be achieved. 
While individual interactions between any two constituents can be weak,
a single constituent might still experience strong collective effects. 
These effects are non-perturbative in nature
and cannot be captured within a perturbative framework. Shifman, Vainshtein and Zakharov suggested in 
\cite{Shifman, Shifman1} to map these non-perturbative effects to the physics of constituents immersed 
in a mean field. The mean field can be expanded in terms of certain condensates. These condensates 
originate as normal-ordered operator products in the standard Wick expansion, which are not required 
to vanish in the ground state. These condensates should be regarded as non-trivial background sources
creating an effective potential to which individual constituents are sensitive. 
In other words, since the bound state, which in the semi-classical limit should be described by 
a classical background metric,
can be viewed as a relevant deformation of the perturbative Minkowski vacuum at the quantum level, 
condensation with respect to that vacuum can be expected to take place.
In such a situation,
normal-ordered products in the Wick expansion have to be taken seriously. Due to these contributions,
observables that are related to $(2N+k)$-correlation function are non-trivial even for $2N>k$.


\section{Parton-level results}\label{plr}	
This section analyses in detail the following situation:
We consider the model of a neutral scalar field described by a hermitian operator $h(x)$.
For simplicity we assume that there are just two eigenvalues of $\Box$, zero and $M^2$,
and that $\langle\Omega|h(x)|q\rangle\not\equiv 0$ for $\Box |q\rangle=0$, but
$\langle\Omega|h(x)|\mathcal{B}\rangle\equiv 0$, $ \langle\Omega|h(x_1)\cdots h(x_N)|\mathcal{B}\rangle\not\equiv 0$
if $\Box |\mathcal{B}\rangle=M^2 |\mathcal{B}\rangle$. In addition, we assume that both states have spin zero. 
Hence, the spectrum of the theory is assumed to consist of two objects, one elementary particle 
which is massless and a massive bound state which is composed of these elementary particles.

Note that at the parton-level the difference between the distribution of scalar and graviton constituents 
is only encoded in a numerical pre-factor, which can be seen as follows: Consider the gauge invariant
auxiliary current $\mathcal{J}(x) = (\Pi^{\mu\nu}h_{\mu\nu})^N(x)$, where 
$\Pi^{\mu\nu}\equiv \eta^{\mu\nu}-\partial^\mu\Box^{-1}\partial^\nu$ denote the components 
of the transverse projection operator. Choosing the harmonic gauge, 
$\partial_\lambda h^\lambda_{\; \mu} = \partial_\mu h^\lambda_{\; \lambda}/2$, the auxiliary current 
reduces to $\mathcal{J}(x)=(h^\lambda_{\; \lambda}/2)^N$. In the auxiliary current description, 
the neutral scalar field introduced above is simply given by $h=h^\lambda_{\; \lambda}/2$.
This current is perfectly consistent with the macroscopic description of a Schwarzschild black-hole.
Indeed, since such a black hole is non-rotating, this feature must be realised quantum mechanically
in such a way that the auxiliary current has spin-$0$. This is obviously the case 
for the choice $(h^\lambda_{\; \lambda}/2)^N$.
 
Moreover, the graviton propagator becomes $\Delta=G \Delta^{(0)}/2$, where $G$ denotes the
Lorentz-covariant generalisation of the Wheeler-de Witt metric, and $\Delta^{(0)}$ is the 
propagator of a free scalar field. When evaluating observables, contractions involving $G$
lead to numerical pre-factors that are inconsequential for the main results of this article
(e.g.~scaling relations for observables).
Thus, at the parton-level, it suffices to work with a massless neutral scalar field $h$.
Since we are dealing with structural properties of black holes
at the parton-level, 
peculiarities of the graviton self-interaction due to a non-polynomial action are of no 
concern for this study. 
Non-perturbative effects due to collective potentials 
experienced by individual gravitons, however, will be taken into account, assuming that individual 
graviton-graviton interactions are weak inside the black hole. 

In order to describe the bound state we introduce the multi-local auxiliary current 
\begin{eqnarray}
	\mathcal{J}(x,\zeta)
	&=&
	{\rm T} h(x+\zeta_1)\cdots h(x+\zeta_N) \;, \hspace{0.5cm} \sum_{a=1}^N \zeta_a =0 \; .
\end{eqnarray}	
Following the arguments presented in Section \ref{acd}, an appropriate local auxiliary current 
is then given by
\begin{eqnarray}
	\mathcal{J}(x)
	&=&
	(2\pi)^{-3/2} \Gamma^{-1}
	\lim_{\zeta\rightarrow 0}  
	{\rm T} h(x+\zeta_1)\cdots h(x+\zeta_N) 
	\; ,
\end{eqnarray}	
with the current normalisation $\Gamma\equiv \langle\Omega|h^N(0)|\mathcal{B}\rangle$.
Furthermore, $\langle\Omega|\mathcal{J}(x)|\Omega\rangle=0$ is implicitly 
assumed. The latter can be realised by subtracting $\langle\Omega|\mathcal{J}(x)|\Omega\rangle$
from $\mathcal{J}(x)$.

Let us first calculate the current normalisation. 
Using the auxiliary current description to represent $| \mathcal{B} \rangle$, the normalization condition $\langle \mathcal{B} | \mathcal{B}\rangle = 1$ becomes
\begin{eqnarray}
\label{normalization}
\Gamma^2 = \int {\rm d}^3 k {\rm d}^3 p \; \mathcal{B}^*(k) 
\mathcal{B}(p) 
\int {\rm d}^3x {\rm d}^3y\; \tfrac{e^{-{\rm i}k\cdot x}}{(2\pi)^{3/2}} \tfrac{e^{{\rm i}p\cdot y}}{(2\pi)^{3/2}}
\langle \Omega |
\mathcal{J} (x) \mathcal{J} (y) | \Omega \rangle  \; \delta_{\mathcal{B}}.
\end{eqnarray}
Here, $\delta_{\mathcal{B}}$ indicates that we are considering correlations at equal time\footnote{
Alternatively, we could work with time-ordered products at different times and use solely covariant integration measures.
The on-shell condition is then understood implicitly and is realised as usual when performing the integration over
the zero components of the momenta. The results are, of course, unaffected.}.
In turn, we can evaluate (\ref{normalization})  using Wick's theorem.
It can be shown that all possible loops can be reduced to self-loops in the limit of large black-hole masses.
We go through this exercise when calculating the distribution function. 
As explained in the section on composite-operator renormalisation in free field theories, 
all such contributions can be safely set to zero at the parton level.
The only non-trivial connected diagram is the one where 
a graviton is emitted at ${\bf x}$ and subsequently absorbed at ${\bf y}$,
while all other fields condense.
Thus, the expectation value in (\ref{normalization}) reduces to
$N^2 \Delta(x - y)\langle \Omega | \hspace{-0.1cm}:h^{N-1}(x) h^{N-1}(y) 
\hspace{-0.1cm}:| \Omega \rangle$. 
Fourier-transforming the propagator, we can shift the integration variable 
$q_0 \rightarrow q_0 + p_0 \sim M^2$, where in the last step the on-shell condition 
and the limit $M \rightarrow \infty$ have been used.
The remaining integrations can now be performed trivially. 
This gives rise to a contact contribution, i.e.~the condensate becomes local. 
Using translational invariance, we can shift the condensate to the origin.
As a result, we find (\ref{Gamma2}),
\begin{eqnarray}
	\Gamma^2
	&=&
	\left(\tfrac{N}{M}\right)^2 \langle\Omega|\hspace{-0.1cm}:\hspace{-0.1cm}h^{2(N-1)}(0)\hspace{-0.1cm}:
	\hspace{-0.1cm}|\Omega\rangle
	\int {\rm d}^3p \; \left|\mathcal{B}(p)\right|^2 
	\; \delta_\mathcal{B} \; .
\end{eqnarray}


We proceed with the calculation of the constituent distribution $\mathcal{D}(x)$ within a composite object 
described by the local auxiliary current $\mathcal{J}(x)$. This amounts to calculating 
\begin{eqnarray}
	\label{Dconst}
	\mathcal{D}(x)
	&=&
	 \int {\rm d}^3 p\; \left|\mathcal{B}(p)\right|^2 \mathcal{A}(p,x) \; , \nonumber \\
	\mathcal{A}(p,x) 
	&=&
	\int_\Sigma {\rm d}^3z_1  {\rm d}^3z_2\; {\rm e}^{-{\rm i}p\cdot (z_1-z_2)}
	\langle \Omega|{\rm T}\mathcal{J}(z_1) \mathcal{O}(x,0) \mathcal{J}(z_2)|\Omega\rangle \; ,
\end{eqnarray}	
with $\mathcal{O}(x,0)=h(x)h(0)$. This bi-local operator is anchored in the hypersurface 
$\Sigma=\{P:y(P)=(0,{\bf y}) \}$. 
The four-point correlation function in $\mathcal{A}$
can only be nontrivial if the auxiliary currents are localised on $\Sigma$. Hence,
$\mathcal{A}=\mathcal{A}_\Sigma \delta_\Sigma$, and correspondingly 
for the constituent distribution $\mathcal{D}=\mathcal{D}_\Sigma \delta_\Sigma$,
where $\delta_\Sigma$ indicates that all fields are localised on the spatial hypersurface $\Sigma$.

\begin{figure}[h]
\centering
\includegraphics[scale=0.1]{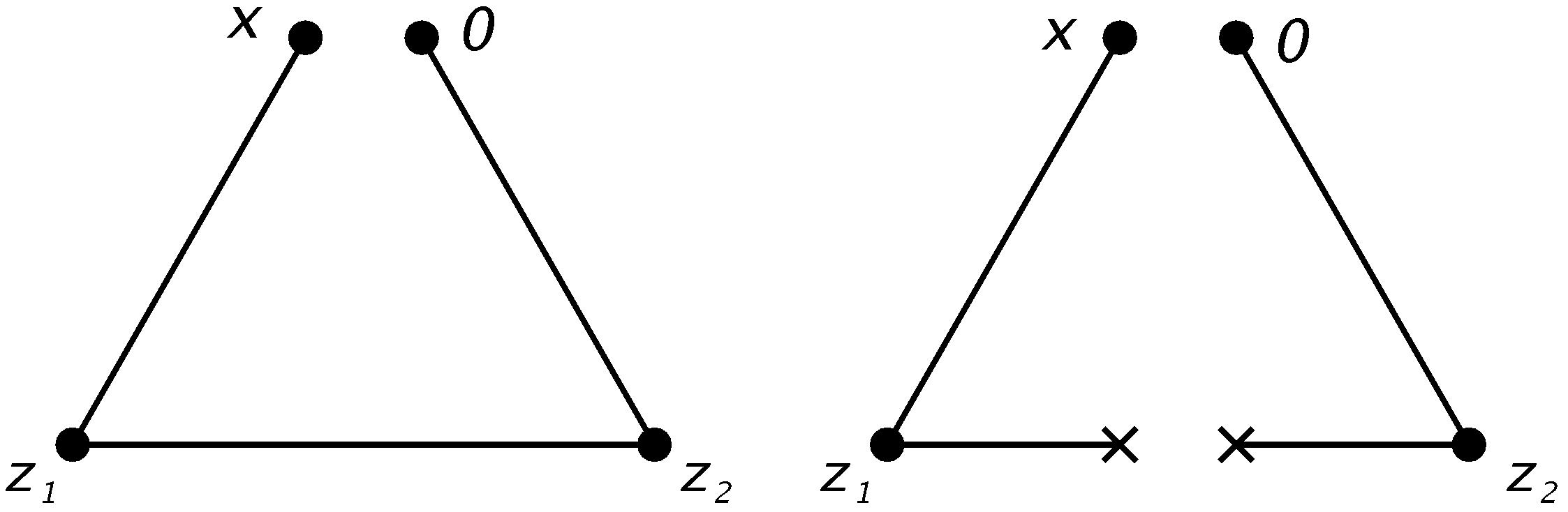}
\caption{Diagrams contributing to the calculation of $\mathcal{D}(r)$ for $N=2$.
The first diagram represents a purely perturbative configuration. 
The second diagram depicts a condensation of space-time events originally located
at $z_1$ and $z_2$. For $N=2$, however, this diagram is disconnected.}
\label{phi2thicka}
\end{figure}

A connected component in 
${\rm T} \mathcal{J}(z_1) \mathcal{O}^{(0)}(x,0)\mathcal{J}(z_2)$ requires $N\ge 2$.
Before considering $N\gg 1$, it is instructive to calculate the minimal connected component
corresponding to $N=2$. This is a purely perturbative contribution. 
Wick expansion of the four-point correlation function 
gives the Feynman diagram shown on the left of Figure~\ref{phi2thicka} (plus a term with $x$ and $0$ exchanged).
We are interested in the limit $M/\mu\rightarrow\infty$, where $\mu$ denotes
any other quantity of mass dimension one. This limit corresponds to a contact
configuration of the two auxiliary currents. Including the term with $x$ and $0$ exchanged,
and using (\ref{Gamma2}), we find 
\begin{eqnarray}
	\label{D0S}
	 \mathcal{D}_\Sigma^{[0]}(x)\Big|_{N=2}
	= 
	\frac{2}{(2\pi)^5}\; \frac{1}{M^{2}} 
	\frac{1}{\langle\Omega|\hspace{-0.1cm}:\hspace{-0.1cm}h^2(0)\hspace{-0.1cm}:
	\hspace{-0.1cm}|\Omega\rangle}
	\; \frac{1}{|{\bf x}|^2}
	\; .  
\end{eqnarray}
Here, $ \mathcal{D}^{[0]}(x)$ denotes the purely perturbative contribution to the constituent distribution, 
which in the context of a free theory refers to the absence of condensates. 
In other words,
condensates in the perturbative contribution only appear in the denominator via the normalisation of $\Gamma$.
In contrast, non-perturbative processes also generate condensates in the numerator (see below).
The constituent distribution as a function of wavelength is found by Fourier-transforming  (\ref{D0S})
and setting $\lambda\equiv \sqrt{2\pi}/|{\bf k}|$,
\begin{eqnarray}
	\mathcal{D}_\Sigma^{[0]}(\lambda)\Big|_{N=2}
	=
	\frac{1}{(2\pi)^5}\; \frac{1}{M^{2}}
	 \frac{1}{\langle\Omega|\hspace{-0.1cm}:\hspace{-0.1cm}h^2(0)\hspace{-0.1cm}:
	\hspace{-0.1cm}|\Omega\rangle} \; \lambda 
	\; .
\end{eqnarray}	
As a result, we find that the constituent distribution depends linearly on the wavelength. In other words,
the bound state $|\mathcal{B}\rangle$ is predominantly populated with soft gravitons.

\begin{figure}[h]
\centering
\includegraphics[scale=0.1]{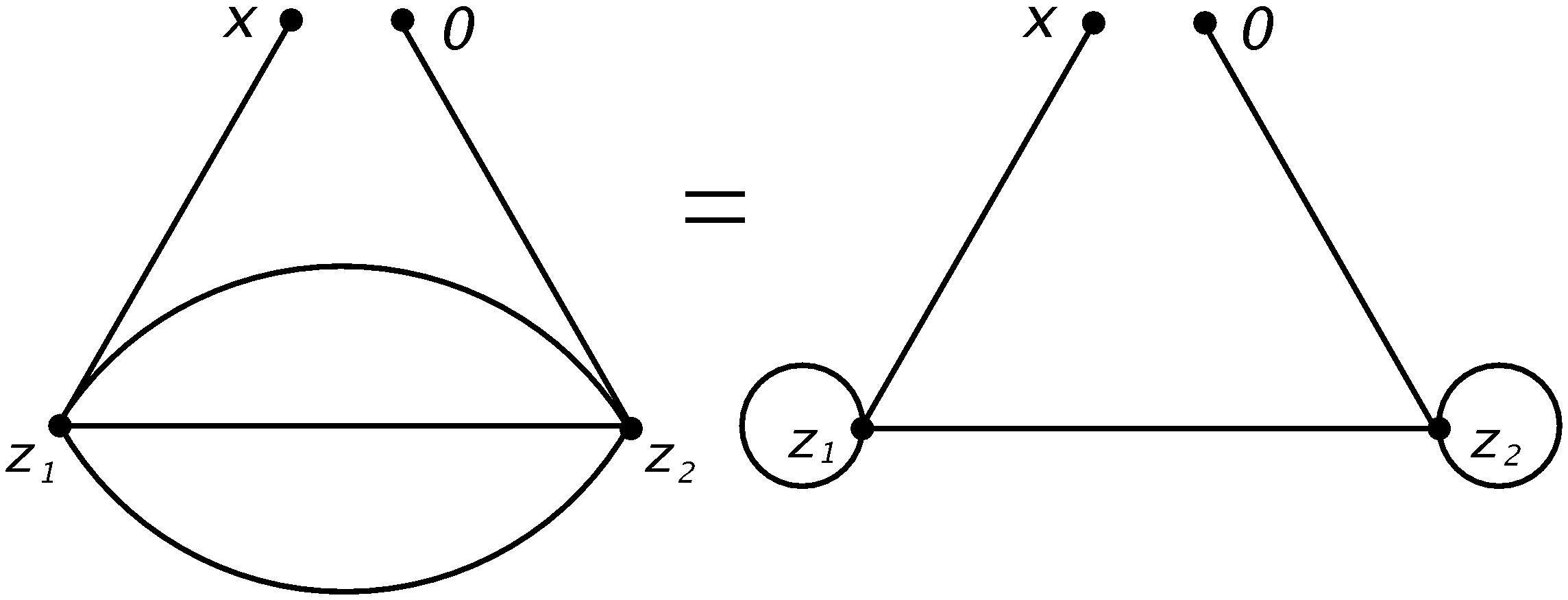}
\caption{Purely perturbative contribution for $N=4$.
In the limit of large black-hole masses, both diagrams 
reduce to the same divergency class. 
Performing composite operator renormalization, these diagrams can be set to zero.}
\label{phi4thickl}
\end{figure}

\begin{figure}[h]
\centering
\includegraphics[scale=0.093]{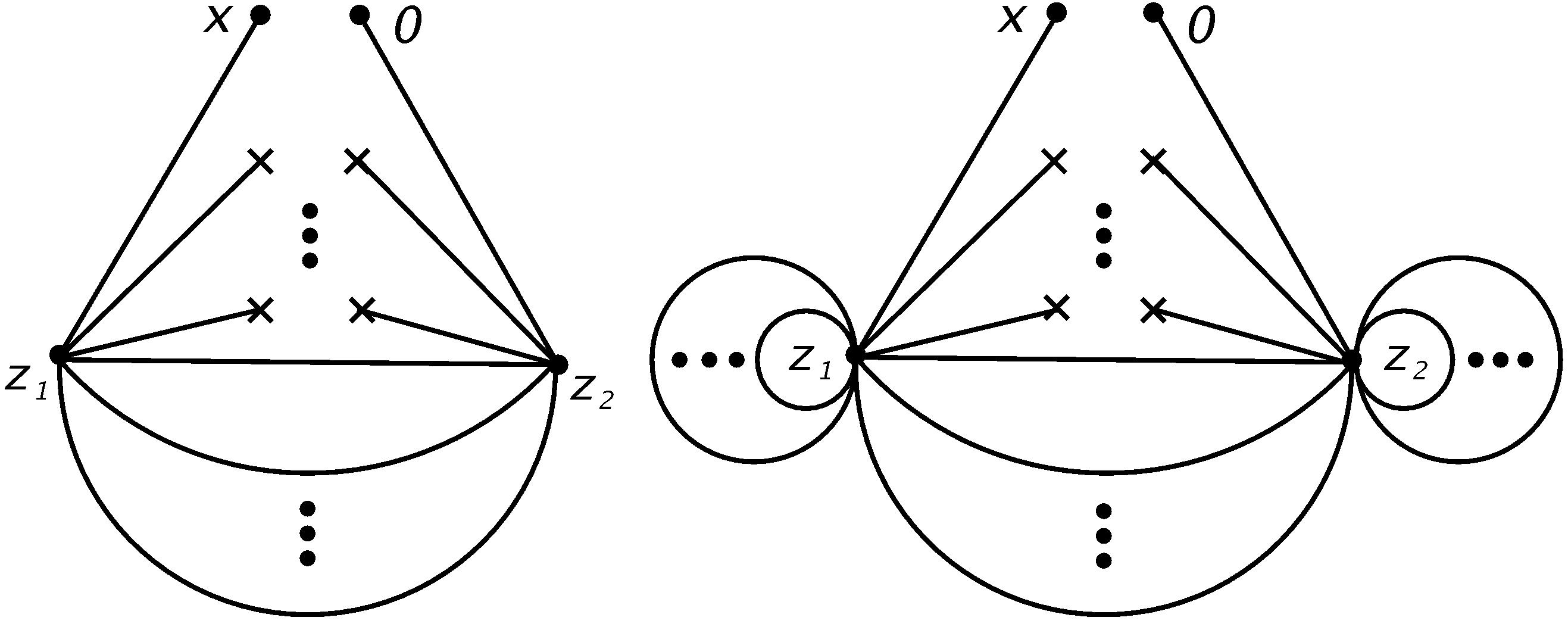}
\caption{The generic situation corresponding to Figure \ref{phi4thickl} for $N$ scalar fields
constituting the auxiliary current. Shown are $k$ condensate insertions and
$l=N-k-2\neq 0$ loops connecting the space--time points $x$ and $0$.
All diagrams with $l>0$ vanish in the limit of large black-hole masses
due to composite operator renormalization.}
\label{phingenericthickl}
\end{figure}

An important question arising from the $N=2$ case is whether 
a purely perturbative contribution is generic for $N\gg 1$. The answer is no.
If the connectivity
between the space--time events at $x$ and $0$ is increased by means of perturbative correlations
(as opposed to condensation), then 
a contribution proportional to $\Delta^{(0)}(0)$ is inevitable in the limit $M/\mu \rightarrow \infty$,
corresponding to a loop anchored at one of the auxiliary currents space--time location, 
as shown in Figures \ref{phi4thickl} and \ref{phingenericthickl}. 
The occurrence of self-loops can be understood as follows. Consider a diagram 
with one loop connecting the positions of the auxiliary currents. This will generate a 
contribution of the form 
\begin{eqnarray}
	&&\int{\rm d}^4 x\; {\rm e}^{-{\rm i}p\cdot x}f(x) \Delta^{(0)}(x)  \Delta^{(0)}(-x)
	\nonumber \\
	 &&= \int \tfrac{{\rm d}^4 k}{(2\pi)^4}   \; f(k) \int \tfrac{{\rm d}^4 k}{(2\pi)^4} \; \Delta^{(0)}(q)
	 \Delta^{(0)}(p-k-q) \nonumber \\
	 &&\stackrel{M/\mu\rightarrow\infty}{\longrightarrow}
	 M^{-2} \int \tfrac{{\rm d}^4 k}{(2\pi)^4}   \; f(k) \int \tfrac{{\rm d}^4 q}{(2\pi)^4} \; \Delta^{(0)}(q) \; .
\end{eqnarray}	
where $p$ denotes the on-shell momentum of the black hole, $p^2=-M^2$, and $f$ is a generic diagram
connected to the loop, which results from Wick expanding the four-point correlation function in the 
definition of $\mathcal{A}(p,r)$. For simplicity we have suppressed all arguments of $f$ irrelevant for 
our discussion.  Thus, the limit $M/\mu\rightarrow\infty$ results in an analytic structure of the diagrams
that is indistinguishable from self-loops. As discussed in Section~\ref{cor}, a proper 
renormalisation prescription at the parton level amounts to setting these contributions to zero. 

\begin{figure}[t]
\centering
\includegraphics[scale=0.11]{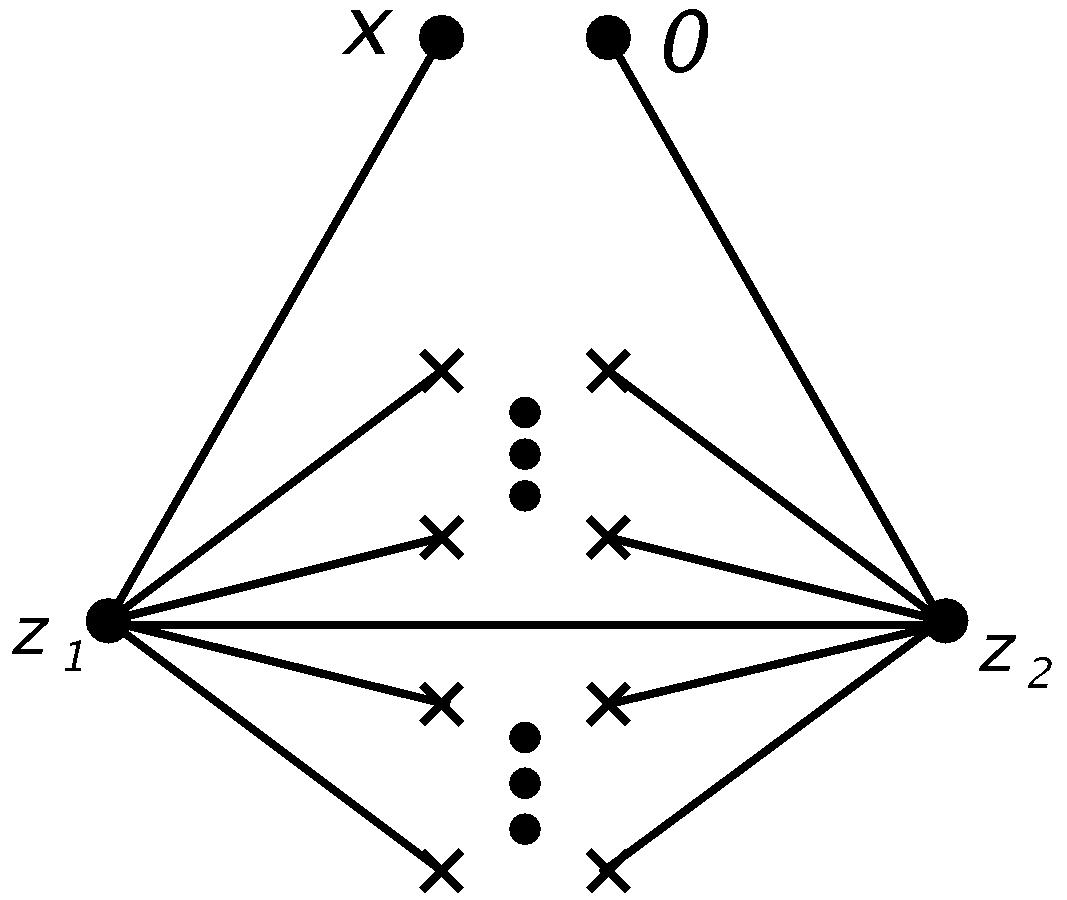}
\caption{Typical diagram contributing to the constituent distribution of a bound state
described by a local auxiliary current composed of $N\gg 1$ constituent fields. 
The four-point correlation function corresponds always to a minimal connected diagram. 
All remaining constituents end up in condensates that parametrise the background 
in which the perturbative degrees of freedom propagate. 
}
\label{phindiagramsthickl}
\end{figure}

As a consequence, even in the general $N>2$ cases, the connected component of $\mathcal{A}(p,x)$
is always minimally connected, i.e.~the number of $h$-propagators is exactly the same as in the 
purely perturbative case for $N=2$. For arbitrary $N>2$, the standard Wick expansion of $\mathcal{A}(p,x)$
corresponds to the diagram shown in Figure~\ref{phindiagramsthickl} plus a diagram 
with $z_1\leftrightarrow z_2$. We find
\begin{eqnarray}
	\label{Apx}
	\mathcal{A}(p,x)
	&=&
	(-{\rm i})^3 \Gamma^{-2} \tbinom{N}{2}^2 \int_\Sigma {\rm d}^3z_1 {\rm d}^3z_2 \;
	\tfrac{{\rm e}^{-{\rm i}p\cdot(z_1-z_2)}}{(2\pi)^3}
	\langle\Omega|\hspace{-0.1cm}:\hspace{-0.1cm}h^{N-2}(z_1)h^{N-2}(z_2)
	\hspace{-0.1cm}:\hspace{-0.1cm}|\Omega\rangle
	\nonumber \\
	&&
	\hspace{3cm} \Delta(x-z_1)\Delta (z_1-z_2) \Delta(z_2) 
\end{eqnarray}	
plus the exchange diagram. 
Inserting a complete set of momentum eigenstates $|k\rangle$ in the condensate, and Fourier-transforming
the propagators, the integrals over the spatial positions of the auxiliary currents can be performed
resulting in the momentum constraints: ${\bf q_2}={\bf p-k+q_1}$, and ${\bf -q_3}={\bf p-k+q_3}$,
where $q_1$ is the four-momentum associated with $x-z_1$, $q_2$ with $z_1-z_2$ and $q_3$
is associated with $z_2$. Shifting the energies of the propagators connecting the observable $\mathcal{O}(x,0)$
with the auxiliary currents, $(q_2)_0\rightarrow (q_2)_0+(p-k)_0$ and $(q_3)_0\rightarrow p_0$,
and taking the limit $M/\mu\rightarrow\infty$, we find (including the exchange diagram)
\begin{eqnarray}
	\mathcal{A}({\bf x})
	&=&
	\frac{2}{(2\pi)^5} \; \frac{\tbinom{N}{2}^2}{M^4} \;   
	\frac{\langle\Omega|\hspace{-0.1cm}:\hspace{-0.1cm}h^{2(N-2)}(0)\hspace{-0.1cm}:
	\hspace{-0.1cm}|\Omega\rangle}
	{\Gamma^2}	
	\; 
	\frac{1}{|{\bf x}|^{2}} \; \delta_{\mathcal{B}} \delta_{\Sigma}
	\; .
\end{eqnarray}	
Fourier-transforming $\mathcal{A}(p,x)$ with respect to the difference vector ${\bf x}$ which connects 
the fields composing $\mathcal{O}(x,0)$ in the hypersurface $x_0=0$, the constituent distribution 
as a function of wavelength is given by
\begin{eqnarray}
	\label{apr}
	\mathcal{D}_\Sigma(\lambda)
	&=& 
	\frac{1}{(2\pi)^5} \frac{(N-1)^2}{M^2} 
	\frac{\langle\Omega|\hspace{-0.1cm}:\hspace{-0.1cm}h^{2(N-2)}(0)\hspace{-0.1cm}:
	\hspace{-0.1cm}|\Omega\rangle}
	{\langle\Omega|\hspace{-0.1cm}:\hspace{-0.1cm}h^{2(N-1)}(0)\hspace{-0.1cm}:
	\hspace{-0.1cm}|\Omega\rangle}
	\; \lambda \;,
\end{eqnarray}	
where (\ref{Gamma2}) has been used. 
The limit $M/\mu\rightarrow\infty$ considerably simplifies the 
calculations of correlation functions involving bound states of mass $M$. This raises the 
question whether (\ref{apr}) is trivial. The answer must be no, since there is no reason to expect that this 
distribution should be trivial, in particular for $M/\mu\rightarrow\infty$. But then 
$M$ cannot be independent of $N$. Moreover, $M/\mu/N\rightarrow$ constant in this limit,
which really is a non-triviality condition and the second indication for $M\propto N$.
This conclusion assumes that the condensate ratio appearing in (\ref{apr}) is $N$-independent.
Relaxing from the limit $M/\mu\rightarrow\infty$, it is clear that $1/N$-corrections will be generated.
Alternatively, the result (\ref{apr}) can be presented in terms of $\mathcal{D}$ (\ref{Dconst}). 
Since $\mathcal{D}\propto\delta_\Sigma$,
only ratios of $\mathcal{D}$ 
evaluated at different length scales are sensible quantities. Denoting by $r_{\rm S}$ an arbitrary 
pivot scale, for instance the Schwarzschild radius which then enters as an external quantity, we have
\begin{eqnarray}
	\mathcal{D}(\lambda)
	&=&
	\mathcal{D}(r_{\rm S}) \; \frac{\lambda}{r_{\rm S}}
	\; .
\end{eqnarray}	   



While $N$ counts the number of $h$-fields composing the auxiliary current or, 
equivalently, its mass dimension, the total constituent number $\mathcal{N}_{\rm c}$ is given by
\begin{eqnarray}
	\mathcal{N}_{\rm c}
	&=& 
	\int{\rm d}^3q \; \langle \mathcal{B}|n({\bf q})|\mathcal{B}\rangle \nonumber \\
	&=&
	\int{\rm d}^3q\; 2q^0 \int {\rm d}^3x \; {\rm e}^{{\rm i}q\cdot x} \; \mathcal{D}(x)\Big|_{q^0=|{\bf q}|}
	\; .
\end{eqnarray}	 
Since $\langle\mathcal{B}|n({\bf q})|\mathcal{B}\rangle = |\mathcal{B}({\bf q})|^2$, 
and due to the on-shell condition, the momentum integral should be restricted to
$|{\bf q}|\in [0,M]$. In the limit
$M/\mu\rightarrow\infty$, $ N\rightarrow\infty:$ $M/\mu/N\rightarrow$ constant, we find
\begin{eqnarray}
	\mathcal{N}_{\rm c}
	&=&
	\frac{1}{3\pi^2} \frac{N^2}{M^2}  
	\frac{\langle\Omega|\hspace{-0.1cm}:\hspace{-0.1cm}h^{2(N-2)}(0)\hspace{-0.1cm}:
	\hspace{-0.1cm}|\Omega\rangle}
	{\langle\Omega|\hspace{-0.1cm}:\hspace{-0.1cm}h^{2(N-1)}(0)\hspace{-0.1cm}:
	\hspace{-0.1cm}|\Omega\rangle}\; 
	M^3 \; \delta_\Sigma \; .
\end{eqnarray}	
As was to be expected, the constituent number diverges 
as the mass dimension of the auxiliary current goes to infinity: 
In fact, for $N_1, N_2\gg 1$, $\mathcal{N}_{\rm c}(N_1)/\mathcal{N}_{\rm c}(N_2)\propto(N_1/N_2)^3$.
Note that this result is consistent with our earlier remark concerning 
$N\neq\mathcal{N}_c$, i.e.~$\mathcal{N}_c$ counts the total number of constituents 
including virtual gravitons which in our formalism are accounted for in terms of condensates.

\begin{figure}[h]
\centering
\includegraphics[scale=0.1]{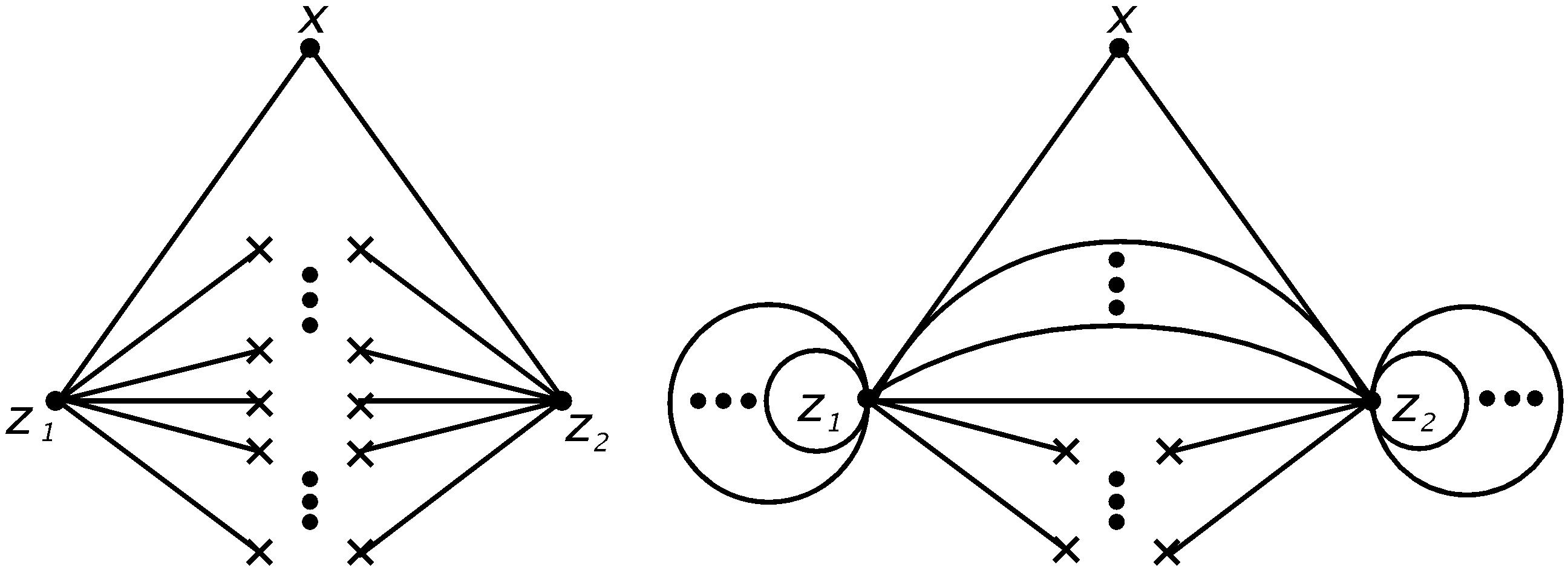}
\caption{Diagrams contributing to the constituents energy-density inside a black hole
represented by a generic auxiliary current. Only the diagram on the left is nontrivial. 
It corresponds to a condensation of all constituent fields not connected to the 
fields composing the observable $\mathcal{E}$.
Increasing the connectivity between the space--time points $z$ and $y$ just by one
propagator 
leads to a vanishing contribution in the limit $M/\mu\rightarrow\infty$
(upon imposing correlation functions to be normal-ordered).}
\label{energydensityphinthickl}
\end{figure}

Given the above scaling behaviour, it is interesting to ask whether the energy density of  
black-hole constituents is a meaningful quantity in the limit 
$M/\mu\rightarrow\infty$, $ N\rightarrow\infty:$ $M/\mu/N\rightarrow$constant. 
At the parton level, it suffices to consider the following energy-momentum tensor:
$\mathcal{T}_{\alpha\beta}=G_{\alpha\beta}^{\; \; \; \; \mu\nu} \partial_\mu h\partial_\nu h/2$,
where $G$ denotes the Lorentz-covariant generalisation of the Wheeler--DeWitt metric.  
Using the auxiliary current description, the standard Wick expansion of 
$\mathcal{E}(x)\equiv \langle\mathcal{B}|\mathcal{T}_{00}(x)|\mathcal{B}\rangle$
results in the type of Feynman diagrams shown in Figure~\ref{energydensityphinthickl}.
By the same reasoning as before when we calculated the distribution function,
all loop corrections vanish in the limit $M/\mu\rightarrow\infty$.
The remaining Feynman diagram is readily calculated to give
\begin{eqnarray}
	\mathcal{E}({\bf x})
	&=&
	\frac{\left|\mathcal{B}({\bf x})\right|^2}
	{2\int {\rm d}^3 p \; \left|\mathcal{B}({\bf p})\right|^2} \; \delta_\Sigma
	\; .
\end{eqnarray}
Both, the total number of constituents as well as the energy density
are defined on the spatial hypersurface $\Sigma$.
In order to define a proper observable, we can consider the energy density per constituent
$\mathcal{E}(x)/\mathcal{N}_{\rm c}$.
While $\mathcal{E}(x)$ and $\mathcal{N}_c$ are temporal distributions
proportional to $\delta_\Sigma$,
the ratio $\mathcal{E}(x)/\mathcal{N}_{\rm c}$ is a physical density
that can be integrated over $\Sigma$ 
to yield the energy per constituent\footnote{
This does not imply that the integral 
of $\mathcal{E}$ over $\Sigma$ is $M$, since $\mathcal{E}\propto\delta_\Sigma$.} 
$\omega=\delta M$, with 
$\delta \equiv c \langle h^{2(N-1)}\rangle/\langle h^{2(N-2)}\rangle/(N M)^2\ll 1$,
where  $c\equiv2/(3\pi^2)$ and $\langle A \rangle\equiv 
\langle\Omega|\hspace{-0.1cm}:\hspace{-0.1cm}A\hspace{-0.1cm}:
\hspace{-0.1cm}|\Omega\rangle$. This is in agreement with our earlier result that 
the black-hole interior is predominantly populated with quanta of the largest possible 
wavelength. Let us introduce the physical constituent number $N_{\rm c}\equiv \delta^{-1}$,
which allows to establish a link between the macroscopic and microscopic description 
of black holes:
\begin{eqnarray}
	\label{M}
	M^{2}
	=
	\frac{2}{3\pi^2} 
	\frac{\langle h^{2(N-1)}\rangle}{\langle h^{2(N-2)}\rangle} \; 
	\frac{N_{\rm c}}{N^2}\;,
\end{eqnarray}
Introducing 
$E^2\equiv 2/(3\pi^2) \langle h^{2(N-1)}\rangle/\langle h^{2(N-2)}\rangle/N^2$, 
the characteristic energy scale $E$ can be related to the typical 
energy per condensed constituent. It
depends on a condensate ratio that is 
a phenomenological input parameter. At this stage, 
at the parton level, we cannot make strong claims about the value 
of this ratio. 

In terms of the characteristic energy scale $E$ we find
\begin{eqnarray}
	M 
	&=&
	\sqrt{N_{\rm c}} \; E
	\; .
\end{eqnarray}	  
This scaling relation shows that the limiting processes 
are self-consistent and capture the correct physics. The consistence and non-triviality 
requirements, as well as simplicity are all granted by the well-established benefits
of field theories with a large number of constituents. 

\section{Outlook: Beyond a partonic description}
\label{bpl}
There are various corrections to the results presented in the last section,
which have been established in the large$-N$ limit of a free field theory. 
First, perturbative graviton exchanges give rise to a series in the 
gravitational coupling strength. Second, non--perturbative contributions 
arise due to strong collective gravitational potentials experience by 
individual constituents. Following the logic of the framework presented 
here, collective effects can be parametrised by condensates, i.e.~for the 
case at hand by curvature condensates (corresponding to field-strength 
squared condensates in Yang-Mills theories). In this section, we sketch
the general strategy for incorporating these corrections in a pragmatic
fashion. Detailed calculations are left for future research. 

Let us demonstrate the appearance of graviton condensates for the case of
gravitational bound states containing scalars $\Phi$ as well as gravitons. For simplicity, 
we assume the scalar to be minimally coupled to gravity and restrict the discussion 
to the distribution function of the scalars. 
Note that at the parton level the contribution coincides 
with the result presented in Section \ref{plr}.
As discussed before, this exercise is not only of academic interest, but also of physical significance.
If a shell of scalar matter collapses, it will source gravity. The resulting state
then consists of both, scalars as well as longitudinal gravitons.
Subsequently, distribution functions for both fields can be defined in accordance with gauge-
invariance. Here we show how the distribution of scalars is affected by gravity.
Note that the construction is reminiscent of quark distribution functions
inside a hadron when interactions are switched on. 
Also there, the distribution of quarks is influenced non-trivially by the presence 
of gauge condensates.
Having this physical situation in mind, let us now discuss our strategy for computing the scalar distribution
in the presence of gravity.

In order to relate gravitons to curvature, the following gauge is useful:
\begin{eqnarray}
	x^\lambda x^\sigma \Gamma^{\mu}_{\lambda\sigma}(x) = 0
	\;, 
\end{eqnarray}
which is the exact analogue of the Fock--Schwinger gauge,
originally proposed in electrodynamics and heavily employed in quantum chromo dynamics. 
In gravity it corresponds to the
choice of a well--known coordinate neighbourhood called a (pseudo-)Riemannian 
normal-coordinate system. Indeed, the Fock--Schwinger gauge is equivalent
to $x^\mu g_{\mu\nu}(x) = x^\mu g_{\mu\nu}(0)$, which in combination with
$g_{\mu\nu}(0)=\eta_{\mu\nu}$ defines a normal coordinate system anchored at $0$.
The geodesic interpretation is that straight lines through the origin parametrize
geodesics in these coordinates.
The Fock--Schwinger gauge allows to conveniently express the potential 
$\mathcal{G}_\mu\equiv \Gamma^\lambda_{\lambda\mu}$ in terms of the Ricci tensor,
\begin{eqnarray}
	\label{grnc}
	\mathcal{G}_\mu(x) &=& -\tfrac{1}{3} x^\lambda R_{\lambda\mu}(0) + \cdots 
	 \; .
\end{eqnarray}	
Terms suppressed in this expansion involve covariant derivatives and products
of Riemann tensors. Although a closed formula for the Riemann normal 
coordinate expansion of $\mathcal{G}(x)$ in local operators can be given, 
it suffices to work with (\ref{grnc}) to illustrate the main idea. 

Consider a $\Phi$--quantum emitted at the space--time point $y$ and absorbed at $x$.
The propagator $\Delta(x,y)\equiv {\rm i}\langle\Omega|{\rm T}\Phi(x)\Phi(y)|\Omega\rangle$
satisfies $(-\Box+\mathcal{G}\cdot\partial)\Delta(x,y)=\delta(x-y)$.
Assuming $\mathcal{G}$ to be small as compared to the free propagation
scale $x-y$, $\Delta(x,y)$ can be expanded as
\begin{eqnarray}
	\label{del}
\Delta(x,y)
	&&=
	\sum_{n=0}^\infty \Delta^{(n)}(x,y) \;, 
	\nonumber \\
		\Delta^{(n)}(x,y)
		&&=
		\int{\rm d}^4z_1\cdots{\rm d}^4z_n \; (-1)^n \Delta^{(0)}(x-z_1) 
		\nonumber \\
		&&\times \mathcal{G}\cdot\partial\Delta^{(0)}(z_n-y)
		\prod_{a=1}^{n-1} \mathcal{G}\cdot\partial
		\Delta^{(0)}(z_a-z_{a+1}),
\end{eqnarray}
where $\Delta^{(0)}$ denotes the free propagator .
This formula has a simple diagrammatic 
interpretation, shown in Figure~\ref{pef}.
The free propagator $\Delta^{(0)}$
transforms invariant under space--time translations, while   
$\Delta$ is non--invariant, since $\mathcal{G}$ depends on the 
space--time location. Given that $\mathcal{G}$ is external and tied
to the ground state properties, this space--time dependence is 
fictitious when the averaged ground state structure is considered. 
There is, however, a second reason for breaking translation invariance.
Namely, once we choose Fock--Schwinger gauge for evaluating (\ref{del}),
the origin of the Riemann normal coordinate neighbourhood is distinguished.
But this is simply due to choosing a coordinate system and bears no 
physical significance, provided all calculations are performed in these
coordinates.
\begin{figure}[h]
\label{pef}
\centering
\includegraphics[scale=0.12]{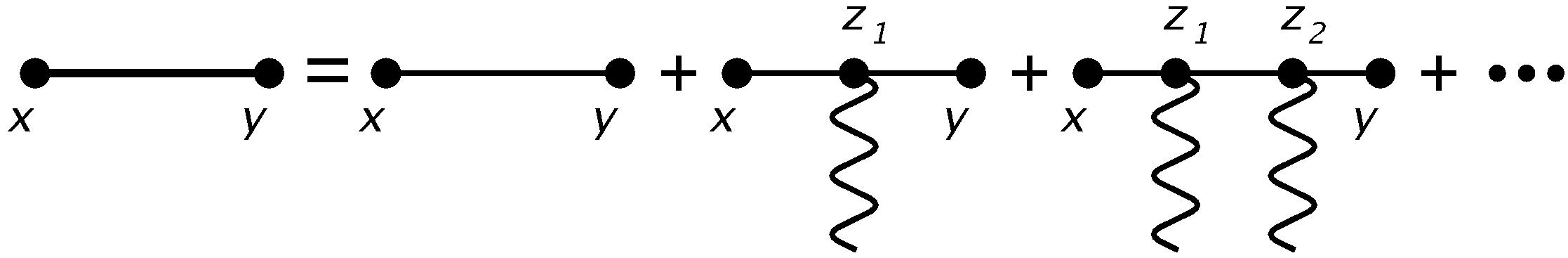}
\caption{Diagrammatic representation of the scalar propagator in the 
external $\mathcal{G}$--field. 
The constituent scatters zero, one, two, $\ldots$ times off the external potential $\mathcal{G}$, 
represented by the wavy lines. On the light--cone, the series of interactions can be summed up, 
resulting in a path--ordered exponential of the 
connection $\mathcal{G}$, in accordance with gauge invariance.}
\label{pef}
\end{figure}  

Having a bookkeeping procedure in mind such as the operator product expansion,
there might be situations where we are only interested in the $R_{\mu\nu}$ contribution.
Then, effectively, $\mathcal{G}_\mu(x)=-x^\lambda R_{\lambda\mu}(0)/3$. 
Other operators in the Riemann normal coordinate expansion of $\mathcal{G}$ 
cannot result in $R$--contributions to $\Delta$.
An elementary calculation using dimensional regularization and the modified minimal subtraction scheme gives
\begin{eqnarray}
	\label{Delta1}
	\Delta^{(1)}(x,y)
	=
	\tfrac{-{\rm i}}{96\pi^2} \; \langle R(0) \rangle
    \left\{
	\ln{\left(\tfrac{y^2}{d^2}\right)} - 1 -
	\tfrac{y^2-(x-y)^2}{(x-y)^2}
	\left[\ln{\left(\tfrac{y^2-(x-y)^2}{y^2}\right)}-1\right]
	\right\}.
\end{eqnarray}
Here, $d$ denotes an arbitrary renormalization length scale. 
Note that (\ref{Delta1}) is exact up to condensates of operators with mass dimensions larger than two, 
which are not shown here.
The Ricci condensate 
$\langle R(0)\rangle\equiv \langle\Omega|\hspace{-0.1cm}:\hspace{-0.1cm}R(0)\hspace{-0.1cm}:
\hspace{-0.1cm}|\Omega\rangle $ 
originates from the condensation of $\mathcal{G}$.
This highlights the practical value of the external field method in Fock-Schwinger gauge
for the non-perturbative description of bound states.

\begin{figure}[h]
\centering
\includegraphics[scale=0.1]{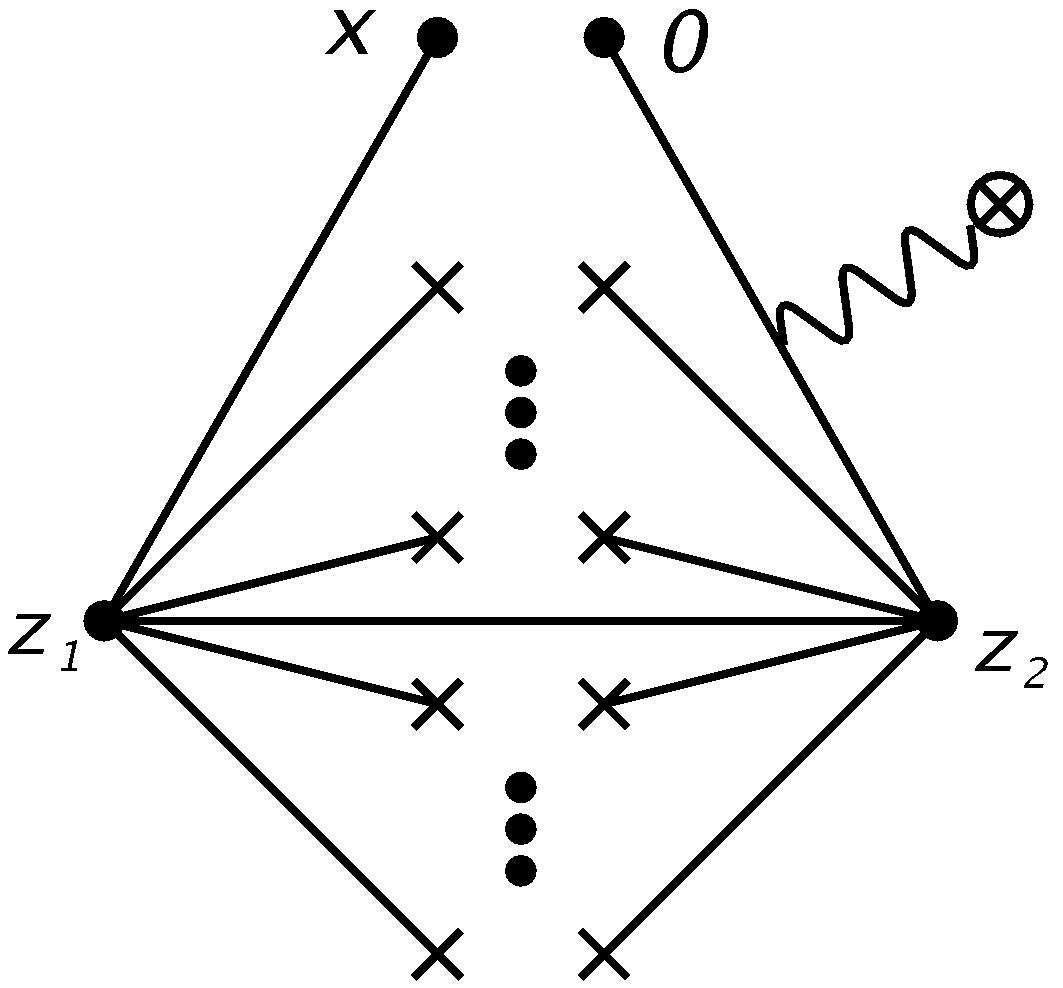}
\caption{An example for a gauge correction to the constituent distribution,
resulting in a Ricci condensate indicated by the sidled line.}
\label{phingaugel}
\end{figure}

As an example for a gauge correction to the distribution of $\Phi$-constituents, 
consider the diagram
shown in Figure \ref{phingaugel}, which gives rise to a contribution proportional to the 
condensate
$\langle \Phi^{2(N-2)} R \rangle\equiv \langle\Omega|\hspace{-0.1cm}:
\Phi^{2(N-2)}\hspace{-0.1cm} R\hspace{-0.1cm}:
\hspace{-0.1cm}|\Omega\rangle$. 
The amplitude is given by (compare to (\ref{Apx}))
\begin{eqnarray}
	\label{ampg}
	\mathcal{A}^{{\rm gc}}(p,r)
	&&=
	\tfrac{1}{96\pi^2 \Gamma^2}\tbinom{N}{2}\int_\Sigma {\rm d}^3x {\rm d}^3y
	\; \tfrac{{\rm e}^{-{\rm i}p\cdot(x+y)}}{(2\pi)^3}
	\Delta^{(0)}(x)\Delta^{(0)}(x-y_-)
	\nonumber \\
	&& \times\left[\ln{\left(\tfrac{-(y_-)^2}{d^2}\right)}-2\right]
	\left\langle \Phi^{N-2}(x) \left(\Phi^{N-2} R\right)(0)\right\rangle,
\end{eqnarray}	
where $y_-:=y-r/2$.
Expanding 
$\Phi^{N-2}(x) (\Phi^{N-2} R)(0)$ into local operators 
yields coefficients suppressed by powers of $p^2=-M^2$.
In the limit $M/\mu\rightarrow \infty$ (with $\mu$ denoting an arbitrary energy scale),
the leading contribution is given by
\begin{eqnarray}
	\mathcal{A}^{{\rm gc}}({\bf p},{\bf r})
	&=&
	\tfrac{1}{96\pi^2\Gamma^2}\tfrac{\tbinom{N}{2}^2}{M^4}
	\; {\rm e}^{{\rm i}{\bf p}\cdot {\bf r}/2} \left[\ln{\left(\tfrac{{\bf r}^2}{{\bf d}^2}\right)}-2\right]
	\left\langle \Phi^{2(N-2)} R\right\rangle \delta_\mathcal{B} \delta_\Sigma
	\; .
\end{eqnarray}	
The corresponding correction to the constituent
distribution is given by 
\begin{eqnarray}
	\mathcal{D}_\Sigma ({\bf r})
	&=&
	\frac{1}{4}\frac{1}{96\pi^2}\frac{N^2}{M^2}\; 
	\frac{\left\langle \Phi^{2(N-2)} R\right\rangle}
	{\left\langle \Phi^{2(N-1)}\right\rangle}
	\frac{\left|\mathcal{B}({\bf r})\right|^2}{\int{\rm d}^3p\; \left|\mathcal{B}({\bf p})\right|^2}
	\left[\ln{\left(\tfrac{{\bf r}^2}{{\bf d}^2}\right)}-2\right]
	\; .
\end{eqnarray}	
This concludes our outlook, which was intended to show how gauge corrections can be incorporated.
As systematic study of the physics of these corrections will be left for further investigations. 

\section{Summary, Conclusions \& Outlook}
\label{sum}
In this article, a description of bound states consisting of $N\gg1$ constituent quanta
has been given. It is based on a relativistic quantum theory of individually weakly coupled 
constituent fields that experience strong binding mechanisms caused by the collection of constituents. 
The quantum states associated with the composite objects are represented by auxiliary currents
composed of the constituent fields in accordance with the quantum numbers and
isometries carried by the bound states. 
This implies the usual reduction formalism pertinent to the asymptotic framework
of scattering theory, but allows also the reduction of kinematical states representing 
bound states beyond an asymptotic framework.  

As an application, Schwarzschild black-holes have been considered as bound states of 
$N\gg1$ constituent gravitons (of all polarisations). Strictly following the logic 
of the framework presented here, we calculated the wave\-length-distribution 
$\mathcal{D}(\lambda)$ 
of constituents inside black holes at the parton level. 
It turns out that the distribution favours to populate black-hole interiors
with constituents of maximal wavelength, 
$\mathcal{D}(\lambda)=\mathcal{D}(r_{\rm S}) \lambda/r_{\rm S}$,
where $r_{\rm S}$ denotes an arbitrary pivot scale, for instance the 
Schwarzschild radius which then enters as an external quantity. 
We showed how gauge corrections arise and how they can be taken into account.
Systematic studies of gauge corrections are, however, left for future publications.
In addition, we calculated the constituent number $\mathcal{N}_{\rm c}$
and the constituents energy density $\mathcal{E}$  
inside the black hole, both of which depend on the hypersurface-localisation.  
Integrating the energy density per constituent $\mathcal{E}/\mathcal{N}_{\rm c}$ 
over a spacial slice, 
we obtained the localisation-independent scaling law
$M=\sqrt{N_{\rm c}} E$ relating the black-hole mass $M$
to the physical constituent number $N_{\rm c}$ and the average energy $E$ per condensed constituent. 
The derivation of this result
is transparent and fully anchored in a field-theoretical context with an interesting interpretation
and relation to previous works such as \cite{tHooft, Witten, Dvali}.

While it is plausible to describe black holes as composite quantum-systems (they certainly allow for
an asymptotic particle-like characterisation), we are convinced that the framework presented here 
allows to illuminate the relation between space-time geometry and quantum physics in general. 
This is left for future work.


\appendix
\section{Constituent density in external fields}
\label{Wilson}
For the sake of a self--contained presentation, in this appendix 
we derive the relation between the bi--local operator $\mathcal{O}$
representing the constituent occupation in the absence and presence of $\mathcal{G}$
to all orders in the derivative coupling on the light--cone. We follow 
\cite{Gross}.

The equation of motion (\ref{eom}) for the diagnostic device $\mathcal{O}$
can be solved iteratively. Including $n$ derivative couplings to the 
gauge connection $\mathcal{G}$, 
the associated bi--local operator at this level is given by
\begin{eqnarray}
	\mathcal{O}^{(n)}(y;r/2)
	&&=
	\int\sigma((z)_n) \; (-1)^n\mathcal{O}^{(0)}(y_+,z_1) 
	\mathcal{G}\cdot\partial\mathcal{O}^{(0)}(z_n,y_-)
	\nonumber \\
	&& \times
	\hspace{0.5cm} \prod_{a\in I(n-1)} \mathcal{G}\cdot\partial
	\mathcal{O}^{(0)}(z_a,z_{a+1})
	.
\end{eqnarray}	
Here, $y_\pm := y \pm r/2$, $I(n)$ denotes the index set $\{1,\dots, n\}$,
$\sigma(z):={\rm d}^4z$ and $\sigma((z)_n):=\sigma(z_1)\cdots\sigma(z_n)$. 

Fourier--transforming the free constituent number operator 
$\mathcal{O}^{(0)}$,
\begin{eqnarray}
	\mathcal{O}^{(n)}(y;r/2)
	&&=
	\frac{(-{\rm i})^n}{(2\pi)^4}\int\sigma(k_0,k_n)\; 
	{\rm e}^{{\rm i}(k_0-k_n)\cdot y}{\rm e}^{{\rm i}(k_0+k_n)\cdot r}
	\nonumber \\ &&\times
\int\sigma((k)_{n-1}) \; F(k_0,(k)_n) \; 
	\prod_{a\in I(n)} k_a\cdot \mathcal{G}(k_{a-1}-k_a)
\end{eqnarray}	
where $F$ denotes the usual propagator denominators for the specified momenta. 
Introducing the new momentum variables $2K:=k_0+k_n\;, 2Q:=k_0-k_n$, 
which are Fourier--conjugated to $y$ and $r$, respectively,
and $q_a:=k_{a-1}-k_a$, gives
\begin{eqnarray}
	\mathcal{O}^{(n)}(y;r/2)
	&&=
	\frac{(-{\rm i})^n}{(2\pi)^4}\int \sigma(K)\sigma(Q)\;
	 {\rm e}^{{\rm i}2Q\cdot y}{\rm e}^{{\rm i}2K\cdot r}
	\nonumber \\
	&& \times \int\sigma((q)_{n}) \delta^{(4)}\Big(Q-\hspace{-0.2cm}\sum_{a\in I(n)}q_a/2\Big) \;
	F\left(K,Q,(q)_{n-1}\right)
	\nonumber \\
	&& \times \prod_{b\in I(n)} \Big(K+Q-\sum_{j=1}^b q_j\Big)\cdot \mathcal{G}(q_b)
	.
\end{eqnarray}
The scale $r$ characterising the diagnostic process is an external scale and can
be further qualified to simplify the expression for $\mathcal{O}^{(n)}(y;r/2)$.
A common qualification is to make it light--like 
and to extract the leading light--cone contribution to $\mathcal{O}^{(n)}(y;r/2)$,
\begin{eqnarray}
	\mathcal{O}^{(n)}(y;r/2)
	&&=
	\frac{(-{\rm i})^nn!}{(2\pi)^4}
	\int{\rm d}u_0 \hspace{-0.2cm}\prod_{a\in I(n)}\hspace{-0.2cm}{\rm d}u_a\; 
	\delta^{(1)}\Big(1-u_0-\hspace{-0.2cm}\sum_{b\in I(n)} u_b\Big)
	\nonumber \\
	&& \times \prod_{c\in I(n)}\int\sigma(q_c) \; 
	\exp{\Bigg\{{\rm i}\sum_{d\in I(n)}q_d \cdot 
	\Big[y-\Big(1-2\sum_{l=1}^du_l\Big)r\Big]\Bigg\}}
	\nonumber \\
	&& \times \int\sigma(P) \; \exp{\left({\rm i}2 r\cdot P\right)} 
	\prod_{m\in I(n)} P\cdot\mathcal{G}(q_c)/(P^2)^{n+1}
	, \nonumber
\end{eqnarray}
where Feynman parameters have been used.
The Fourier--transformation $P\rightarrow r$ requires regularisation. 
Employing the $\overline{\rm MS}$ scheme 
it is readily evaluated:
\begin{eqnarray}
	(2\pi)^4 \frac{{\rm i}^n}{n!} \prod_{a\in I(n)} r^{\lambda_a} \; \mathcal{O}^{(0)}(y;r).
\end{eqnarray}	
Performing the $u_0$--integration, we arrive at
\begin{eqnarray}
	\label{pexp}
	\mathcal{O}(y;r/2)
	=
	\mathcal{P}\exp{\left(-\int\hspace{-0.35cm}C {\rm d}z^\lambda \mathcal{G}_\lambda(z)\right)}
	\; \mathcal{O}^{(0)}(y;r/2),\nonumber
\end{eqnarray}
where $C$ denotes the contour given by the path $z:[0,1]\rightarrow \mathbb{R}^4\;,
u\rightarrow z(u):=y-(1-2 u)r$, and $\mathcal{P}$ refers to path ordering
along this contour.\\
\vspace{0.5cm}

\textbf{Acknowledgements}

\vspace{0.5cm}
It is a great pleasure to thank Lukas Gr\"unding and Florian Niedermann 
for many enlightening discussions and for their constructive criticism 
of earlier versions of this manuscript. 
We thank Kerstin Paech and Florian Niedermann 
for highlightning the relevance of Witten-scaling to us. 
Last but not least we acknowledge discussions with 
Dennis Dietrich, Gia Dvali, Cesar Gomez, Floria K\"uhnel,
Andreas Sch\"afer, Robert Schneider, Bo Sundborg and Nico Wintergerst. 
The work of TR was supported by the Humboldt Foundation and the
International Max Planck Research School on Elementary Particle Physics.
The work of SH was supported by
the DFG cluster of excellence `Origin and Structure of
the Universe' and by TRR 33 `The Dark Universe'






\end{document}